\documentclass{jfm}
\usepackage{graphicx}
\usepackage{epstopdf, epsfig}

\usepackage{amsmath,amssymb} 
\usepackage{tabularx} 
\usepackage{booktabs} 
\usepackage{subfig} 
\usepackage{color}

\usepackage[section]{placeins}

\usepackage[dvipsnames]{xcolor} 
\usepackage{pgfplots} 

\newcommand\Nus{\mbox{\textit{Nu}}}
\newcommand\Sta{\mbox{\textit{St}}}
\newcommand\Mac{\mbox{\textit{Ma}}}

\makeatletter
\AtBeginDocument{%
  \expandafter\renewcommand\expandafter\subsection\expandafter{%
    \expandafter\@fb@secFB\subsection
  }%
}
\makeatother

\newcommand{\pd}[2]{\frac{\partial #1}{\partial #2}} 

\newcommand{\ol}[1]{\overline{#1}}
\newcommand{\wt}[1]{\widetilde{#1}}

\newcommand{\nn}[0]{\nonumber}

\newcommand{\rk}[1]{\left(#1\right)}

\newcommand{\dt}[1]{\pd{}{t}\rk{#1}}
\newcommand{\ub}[1]{\underbrace{#1}}

\definecolor{matc}{RGB}{88,254,249}
\definecolor{matg}{RGB}{128,128,128}

\tikzstyle{solid}=                   [dash pattern=]

\tikzstyle{dotted}=                  [dash pattern=on \pgflinewidth off 2.5pt]

\tikzstyle{dashed}=                  [dash pattern=on 8pt off 5pt]

\tikzstyle{dashdotted}=              [dash pattern=on 6pt off 2pt on 2pt off 2pt]

\tikzstyle{Mdashed}=					[dash pattern=on 7pt off 5pt]
\tikzstyle{Mdashdotted}= 			[dash pattern=on 8pt off 3pt on \the\pgflinewidth off 3pt]

\tikzstyle{Odotted}=          		[dash pattern=on \pgflinewidth off 0.8pt]
\tikzstyle{Odashed}=                 [dash pattern=on 2pt off 0.8pt]
\tikzstyle{Odashdotted}= 			[dash pattern=on 3pt off 2pt on 1pt off 2pt]

\tikzset{mark size=1.2pt}

\newcommand{\budGGW}[0]{{\protect\tikz \protect\draw[red,line width=1.6pt]  (0,0) -- (0.8,0);}}
\newcommand{\budC}[0]{{\protect\tikz \protect\draw[black,line width=0.8pt]  (0,0) -- (0.8,0);}}
\newcommand{\budPR}[0]{{\protect\tikz \protect\draw[black,dotted, line width=0.8pt]  (0,0) -- (0.8,0);}}
\newcommand{\budTD}[0]{{\protect\tikz \protect\draw[black,line width=0.8pt] (0,0) -- (0.8,0) plot[mark=x,mark size=2pt] coordinates {(0.15,0) (0.4,0) (0.65,0)};}}
\newcommand{\budVD}[0]{{\protect\tikz \protect\draw[black,line width=0.8pt] (0,0) -- (0.8,0) plot[mark=*,mark size=1.0pt] coordinates {(0.15,0) (0.4,0) (0.65,0)};}}
\newcommand{\budM}[0]{{\protect\tikz \protect\draw[black,dashed, line width=0.8pt]  (0,0) -- (0.8,0);}}
\newcommand{\budPS}[0]{{\protect\tikz \protect\draw[matg,solid,line width=0.8pt]  (0,0) -- (0.8,0);}}
\newcommand{\budDS}[0]{{\protect\tikz \protect\draw[matg,line width=0.8pt] (0,0) -- (0.8,0) plot[mark=x,mark size=2pt] coordinates {(0.15,0) (0.4,0) (0.65,0)};}}

\newcommand{\link}[0]{{\protect\tikz \protect\draw[black,solid,line width=0.8pt]  (0,0) -- (0.8,0);}}
\newcommand{\linr}[0]{{\protect\tikz \protect\draw[red,solid,line width=0.8pt]  (0,0) -- (0.8,0);}}
\newcommand{\linb}[0]{{\protect\tikz \protect\draw[blue,solid,line width=0.8pt]  (0,0) -- (0.8,0);}}
\newcommand{\linc}[0]{{\protect\tikz \protect\draw[matc,solid,line width=0.8pt]  (0,0) -- (0.8,0);}}

\newcommand{\linpk}[0]{{\protect\tikz \protect\draw[black,dotted, line width=0.8pt]  (0,0) -- (0.8,0);}}

\newcommand{\lindk}[0]{{\protect\tikz \protect\draw[black,dashed, line width=0.8pt]  (0,0) -- (0.8,0);}}
\newcommand{\lindr}[0]{{\protect\tikz \protect\draw[red,dashed, line width=0.8pt]  (0,0) -- (0.8,0);}}

\newcommand{\lineins}[0]{{\protect\tikz \protect\draw[matg,dashdotted,line width=0.8pt]  (0,0) -- (0.8,0);}}
\newcommand{\linzwei}[0]{{\protect\tikz \protect\draw[matg,dashed,line width=0.8pt]  (0,0) -- (0.8,0);}}
\newcommand{\lindrei}[0]{{\protect\tikz \protect\draw[matg,solid,line width=0.8pt]  (0,0) -- (0.8,0);}}
\newcommand{\linvier}[0]{{\protect\tikz \protect\draw[black,dashdotted,line width=0.8pt]  (0,0) -- (0.8,0);}}
\newcommand{\linfuenf}[0]{{\protect\tikz \protect\draw[black,dashed,line width=0.8pt]  (0,0) -- (0.8,0);}}
\newcommand{\linsechs}[0]{{\protect\tikz \protect\draw[black,solid,line width=0.8pt]  (0,0) -- (0.8,0);}}

\newcommand{\linOpk}[0]{{\protect\tikz \protect\draw[black,Odotted, line width=0.8pt]  (0,0) -- (0.8,0);}}
\newcommand{\linOdk}[0]{{\protect\tikz \protect\draw[black,Odashed, line width=0.8pt]  (0,0) -- (0.8,0);}}
\newcommand{\linOddk}[0]{{\protect\tikz \protect\draw[black,Odashdotted, line width=0.8pt]  (0,0) -- (0.8,0);}}

\shorttitle{Statistics of fully turbulent impinging jets}
\shortauthor{R. Wilke and J. Sesterhenn}

\title{Statistics of fully turbulent impinging jets}

\author{Robert Wilke\aff{1}
 \and J\"orn Sesterhenn\aff{1}
 \corresp{\email{joern.sesterhenn@tu-berlin.de}}}

\affiliation{\aff{1}Institute of Fluid Dynamics and Technical Acoustics, TU Berlin,
M\"uller-Breslau-Str. 12, 10623 Berlin, Germany}

\begin{document}

\maketitle

\begin{abstract}

Direct numerical simulations of sub- and supersonic impinging jets with Reynolds numbers of 3300 and 8000 are carried out to analyse their statistical properties. The influence of the parameters Mach number, Reynolds number and ambient temperature on the mean velocity and temperature fields are studied. For the compressible subsonic cold impinging jets into a heated environment, different Reynolds analogies are assesses. It is shown, that the (original) Reynolds analogy as well as the Chilton Colburn analogy are in good agreement with the DNS data outside the impinging area. The generalised Reynolds analogy (GRA) and the Crocco-Busemann relation are not suited for the estimation of the mean temperature field based on the mean velocity field of impinging jets. Furthermore, the prediction of fluctuating temperatures according to the GRA fails. On the contrary, the linear relation between thermodynamic fluctuations of entropy, density and temperature as suggested by \cite{LechnerSesterhenn2001} can be confirmed for the entire wall jet. The turbulent heat flux and Reynolds stress tensor are analysed and brought into coherence with the primary and secondary ring vortices of the wall jet. Budget terms of the Reynolds stress tensor are given as data base for the improvement of turbulence models.

\end{abstract}


\section{Introduction}

Impinging jets are widely used: for the cooling of hot surfaces such as turbine blades, as rocket engine or vertical and/or short take-off and landing (V/STOL) aircraft aero engine. Therefore impinging jets have been studied for decades. Schematic illustrations of the flow fields as well as distributions of local Nusselt numbers for plenty of different geometrical configurations and Reynolds numbers $\Rey$ can be found in several reviews, such as \cite{WeigandSpring2011} based on experimental and numerical results. Since experiments cannot provide all quantities of the flow spatially and temporally well resolved, the understanding of the turbulent flow field requires simulations. For example, the investigation of the generation of tones and the connection between vortex dynamics and heat transfer require precise simulations.

Most existing publications of numerical nature use either turbulence modelling for the closure of the Reynolds-averaged Navier-Stokes (RANS) equations, e.g. \cite{ZuckermanLior2005}, or large eddy simulation (LES), e.g. \cite{CzieslaBiswas2001}. Almost all available direct numerical simulations (DNS) are either two-dimensional, e.g. \cite{ChungLuo2002}, or do not exhibit an appropriate spatial resolution in the three-dimensional case. For example, \cite{HattoriNagano2004} performed a 2.5 dimensional 'DNS' on 3 million grid points ($\Rey=9120$) on a huge domain $26 \times 2 \times 1.6$ diameters. As a comparison, the simulations described in this article with $\Rey=8000$ are performed with more than one billion grid points. Recent investigations with an appropriate resolution come from \cite{DairayFortune2015}. He conducted a DNS of a round impinging jet with a nozzle to plate distance of $h/D=2$ and focused on the secondary maximum of the heat transfer distribution and on the connection to elongated structures. However compared to free jets, pipe and channel flows little is known about the turbulence of impinging jets. With this article we want to contribute to close the present knowledge gap in literature, as it was done for the pipe and channel flow, e.g. \cite{EggelsUnger1994}, \cite{LechnerSesterhenn2001}.
 
Different utilisations of impinging jets involve different flow conditions: sub- and supersonic impinging jets, jets with ambient temperature or cold jets in a hot environment. In this article we firstly address the influence of those parameters on the mean field of the impinging jet. Due to the huge amount of parameters ($\Rey$, Mach number $\Mac$, cold/hot environment, radial and axial position within the domain), in this article a detailed description of the turbulence statistics is given only for the cold impinging jet in a hot environment ($\Rey=8000$, $\Mac=0.8$). We further concentrate on the wall jet region. The flow close to the nozzle exit is equal to the free jet and therefore not analysed within this article. The presence of the impinging plate influences the jet within the last two diameters adjacent to the wall.

\section{Computational setup}
\label{sec:geometry}

The governing Navier-Stokes equations are formulated in a characteristic pressure-velocity-entropy-formulation ($p,u,v,w,s$), as described by \cite{Sesterhenn2001} and are solved directly numerically. This formulation has advantages in the fields of boundary conditions, space discretisation and parallelization. Since the smallest scales of turbulent motion are resolved, no turbulence modelling is required. The spatial discretisation uses compact 5th order upwind finite differences for the convective terms and 6th order compact central schemes for the diffusive terms. In order to advance in time a 4th order Runge-Kutta scheme is applied. To avoid Gibbs oscillations in the vicinity of the standoff shock (for supersonic impinging jets) an adaptive shock-capturing filter developed by \cite{BogeyCacqueray2009} that automatically detects shocks is used.

\begin{figure}
\centering
\includegraphics[width=\textwidth]{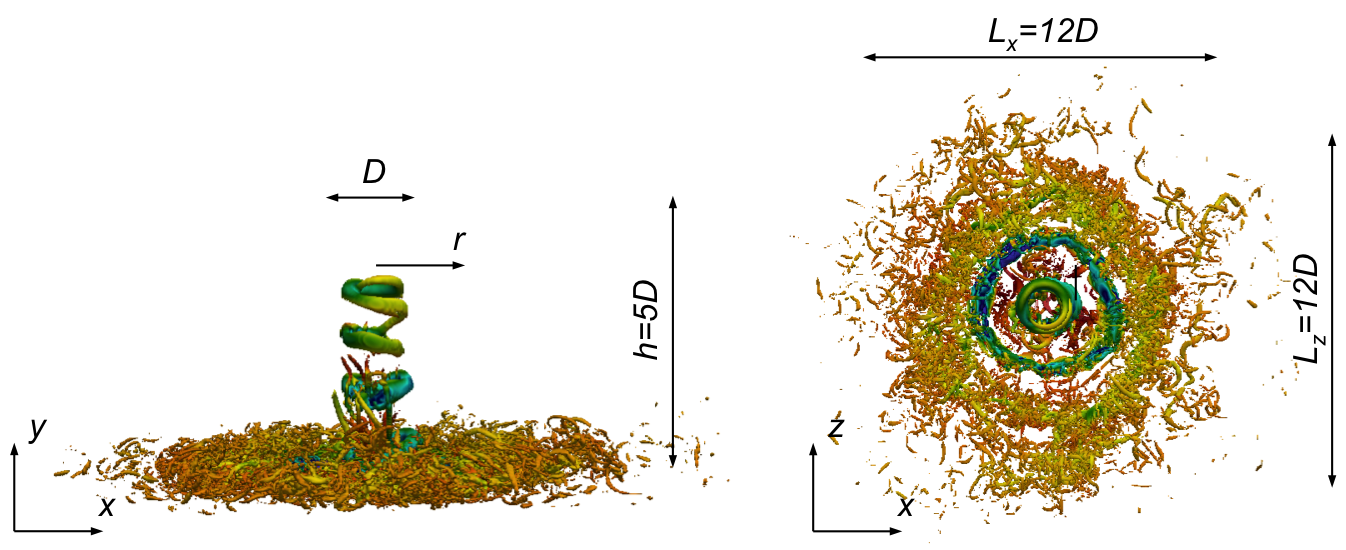}

\caption{3D countour plot of the impinging jet (\#6) at $Q D^2/u_{\infty}^2=5$, coloured with the pressure ($0.8 \leq p/p_{\infty} \leq 1.1$, blue to red)}
\label{fig:domain}
\end{figure}

The computational domain is delimited by an isothermal wall, which is the impinging plate, and one boundary consisting of an isothermal wall and the inlet as well as four non-reflecting boundary conditions. The walls are fully acoustically reflective. The nozzle location is defined using a hyperbolic tangent profile with a disturbed thin laminar annular shear layer as described in \cite{WilkeSesterhenn2014}. A sponge region is applied for the outlet area $r/D >5$, that smoothly forces the values $p,u,v,w,s$ to reference values obtained by a preliminary large eddy simulation of a greater domain. This destroys vortices before leaving the computational domain. The coordinate system and domain sizes are shown in figure \ref{fig:domain}.

The grid is refined in the wall-adjacent regions in order to ensure a maximum value of the dimensionless wall distance $y^+$ of the closest grid point to the wall not larger than one for both plates. For the wall-parallel-directions a slight symmetrical grid stretching is applied, which refines the jet shear layer. The refinements use hyperbolic tangent respectively hyperbolic sin functions resulting in a change of the mesh spacing lower than $1\%$ for all cases and directions. Table \ref{tab_para} shows the physical and geometrical parameters of the simulations.

\begin{table}
	\caption{Geometrical and physical parameters of the simulation. $p_o, p_{\infty}, T_{t,i}, T_{\infty},T_W, \Mac, \Rey, \Pran, \kappa, R$ and $y_W^+$ denote total- and ambient pressure, total-, ambient and wall temperature, Mach, Reynolds and Prandtl number, ratio of specific heats, the specific gas constant and the dimensionless wall distance at the wall.}
	\begin{tabularx}{\columnwidth}{p{6mm} XXXXXXXX p{17mm}}
	\toprule
	N$^{\circ}$ & $p_o/p$ & $p_{\infty}$ & $\Mac$ & $T_{t,i}$ & $T_{\infty}=T_W$ & $Re$ & $Pr$ & $\kappa$ & $R$\\
	 & & [Pa] & & [K] & [K] & && &[J/(kg K)]\\
	\midrule
    \#1 & $2.15$ & $10^5$ & $1.106$ & $293.15$ & $373.15$ & $3300$ & $0.71$ & $1.4$ & $287$\\
    \#2 & $2.15$ & $10^5$ & $1.106$ & $293.15$ & $293.15$ & $3300$ & $0.71$ & $1.4$ & $287$\\
    \#3 & $2.15$ & $10^5$ & $1.106$ & $293.15$ & $293.15$ & $8000$ & $0.71$ & $1.4$ & $287$\\
    \#4 & $1.12$ & $10^5$ & $0.408$ & $293.15$ & $373.15$ & $3300$ & $0.71$ & $1.4$ & $287$\\
    \#5 & $1.50$ & $10^5$ & $0.784$ & $293.15$ & $373.15$ & $3300$ & $0.71$ & $1.4$ & $287$\\
    \#6 & $1.50$ & $10^5$ & $0.784$ & $293.15$ & $373.15$ & $8000$ & $0.71$ & $1.4$ & $287$\\
	\end{tabularx}
	\begin{tabularx}{\columnwidth}{p{6mm} X p{32mm} p{10mm} XX}
	\toprule
	N$^{\circ}$ & domain size & grid points & max. $y_W^+$ & grid width x,z & grid width y\\
	& $[D]$ & & & $[D]$ & $[D]$\\
	\midrule
    \#1 & $12 \times 5 \times 12$ & $512 \times 512 \times 512$ & $0.67$ & $0.0199 .. 0.0588$ & $0.0017 .. 0.0159$\\
    \#2 & $12 \times 5 \times 12$ & $512 \times 512 \times 512$ & $0.77$ & $0.0199 .. 0.0588$ & $0.0017 .. 0.0159$\\
    \#3 & $12 \times 5 \times 12$ & $1024 \times 1024 \times 1024$ & $1.02$ & $0.0099 .. 0.0296$ & $0.0012 .. 0.0072$\\
    
    \#4 & $12 \times 5 \times 12$ & $512 \times 512 \times 512$ & $0.62$ & $0.0184 .. 0.0636$ & $0.0017 .. 0.0159$\\
    \#5 & $12 \times 5 \times 12$ & $512 \times 512 \times 512$ & $0.63$ & $0.0165 .. 0.0388$ & $0.0017 .. 0.0159$\\
    \#6 & $12 \times 5 \times 12$ & $1024 \times 1024 \times 1024$ & $0.58$ & $0.0099 .. 0.0296$ & $0.0008 .. 0.0078$\\
	\bottomrule
	\end{tabularx}
	\label{tab_para}
\end{table}

\section{Mean flow}
\subsection{Wall jet}

According to the radial velocity, we divide the wall jet into four zones. The accelerating zone ($0 \leq r/D \lesssim 0.8$), the zone of maximal radial velocity ($r/D \approx 0.8$), the decelerating zone ($r/D \gtrsim 0.8$) and a zone, where the influence of the impingement is not dominant any more ($r/D \gtrsim 2.5$). According to the parameters, the position of the maximum radial velocity changes slightly. For the flow description, the radial positions $r/D=0.3$, $0.8$, $1.4$ and $3.5$ are chosen.

\begin{figure}
\captionsetup[subfigure]{labelformat=empty}
\centering

\subfloat[]{\includegraphics[width=0.245\textwidth]{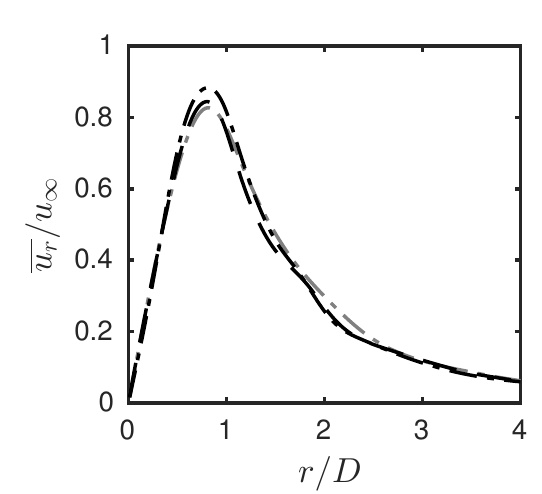}}
\subfloat[]{\includegraphics[width=0.245\textwidth]{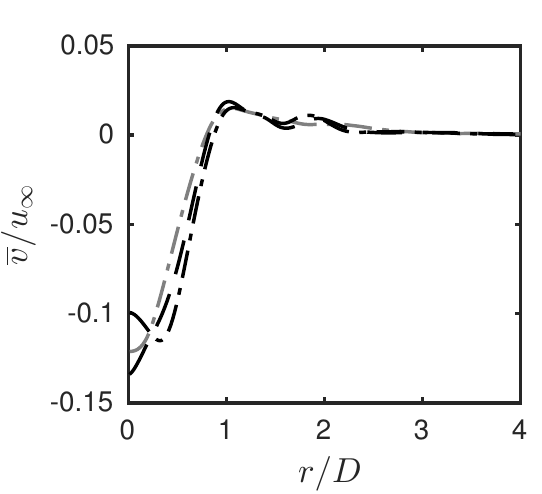}}
\subfloat[]{\includegraphics[width=0.245\textwidth]{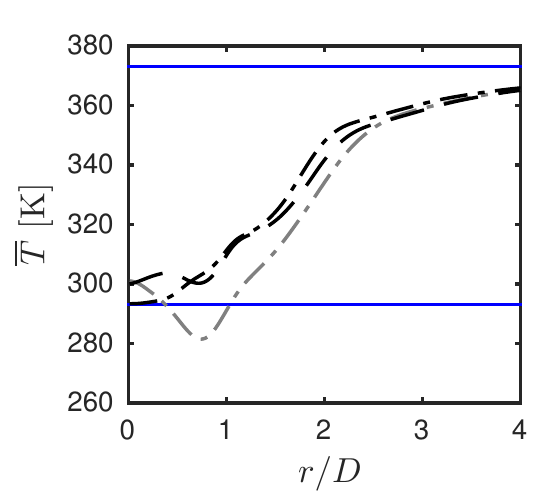}}
\subfloat[]{\includegraphics[width=0.245\textwidth]{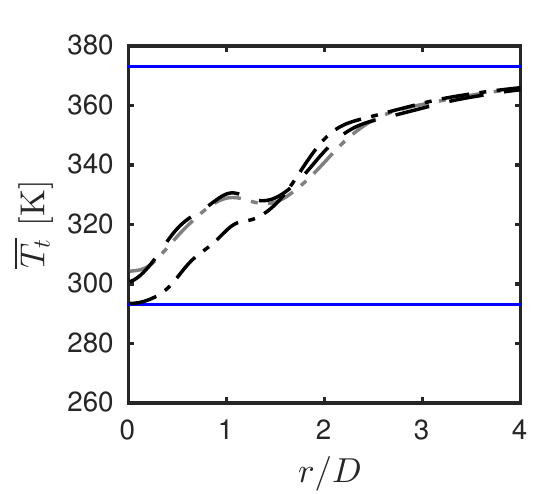}}\\[-8mm]

\subfloat[]{\includegraphics[width=0.245\textwidth]{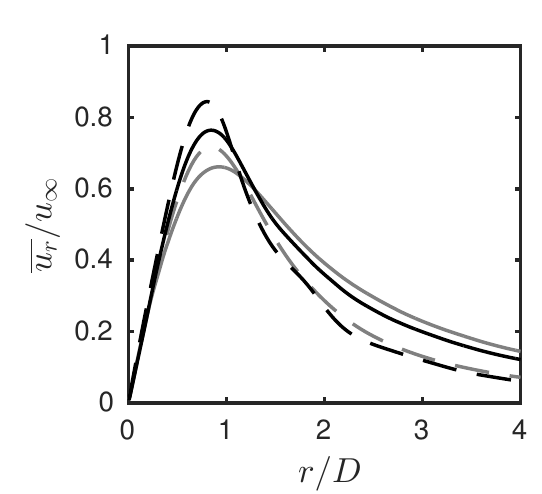}}
\subfloat[]{\includegraphics[width=0.245\textwidth]{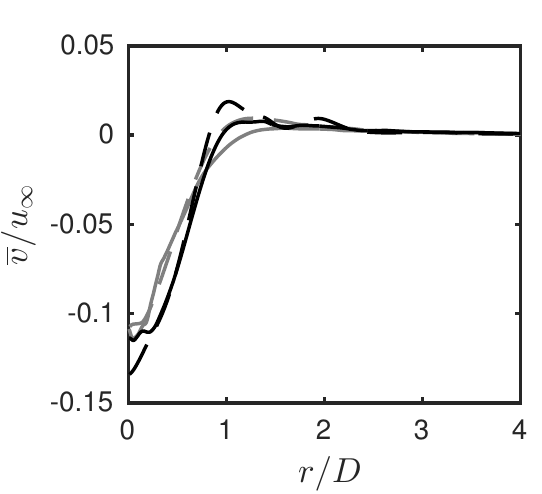}}
\subfloat[]{\includegraphics[width=0.245\textwidth]{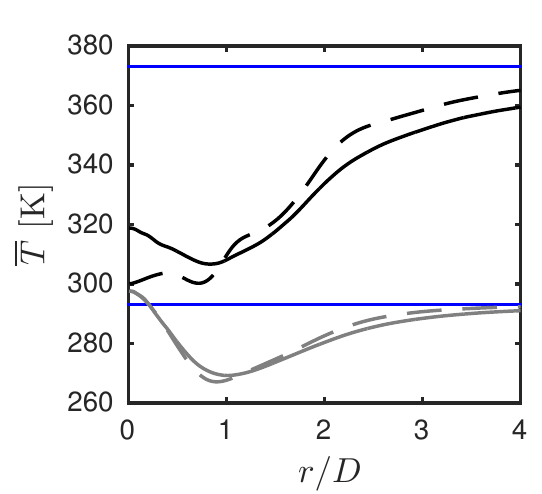}}
\subfloat[]{\includegraphics[width=0.245\textwidth]{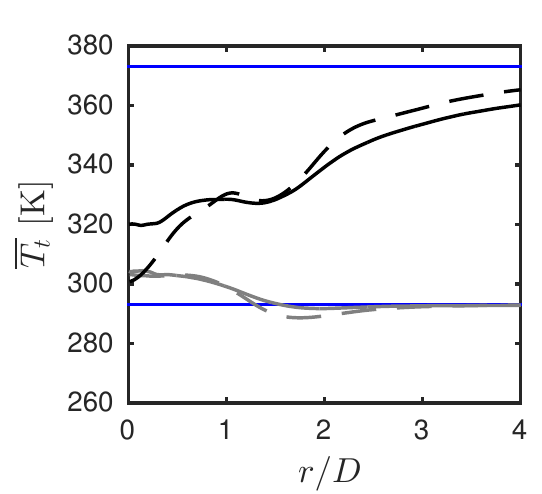}}\\[-8mm]

\subfloat[]{\includegraphics[width=0.245\textwidth]{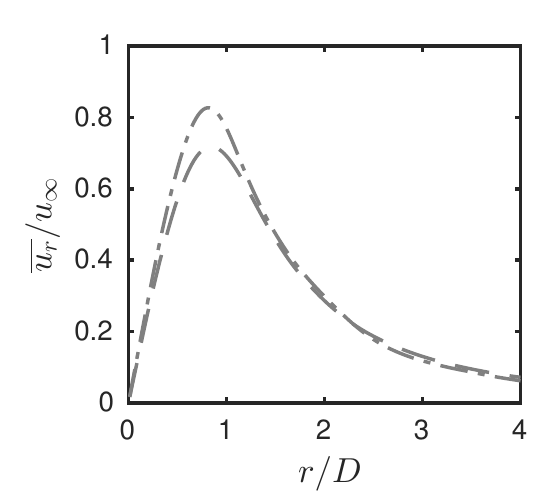}}
\subfloat[]{\includegraphics[width=0.245\textwidth]{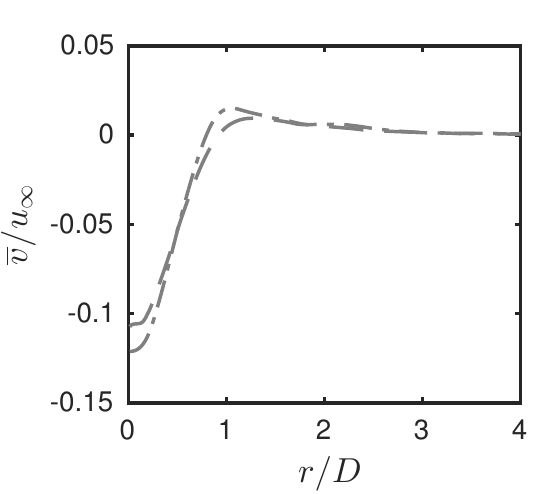}}
\subfloat[]{\includegraphics[width=0.245\textwidth]{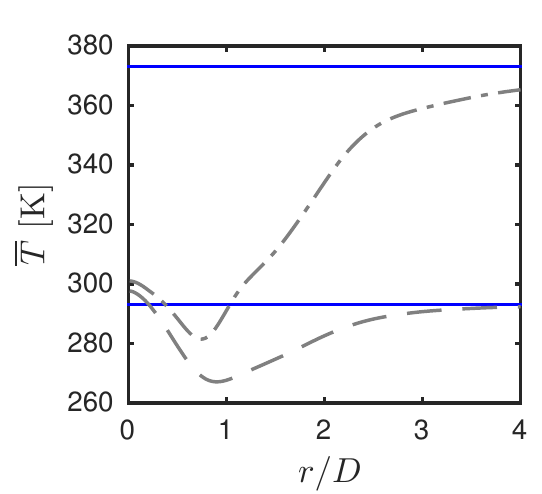}}
\subfloat[]{\includegraphics[width=0.245\textwidth]{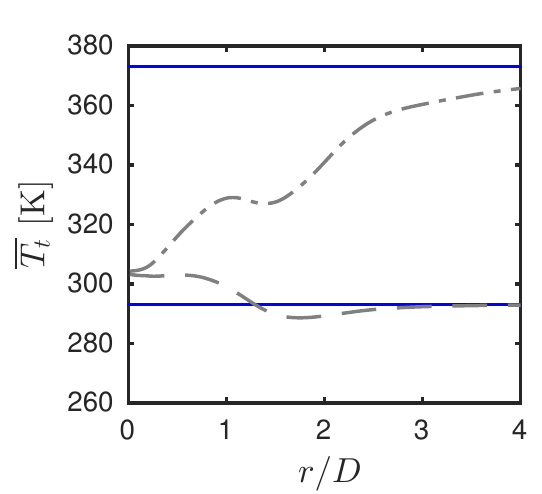}}\\[-6mm]

\caption{Mean profiles (left to right) of the radial velocity $u_r$, axial velocity $v$, temperature $T$ and total temperature $T_t$ at $y/D=0.05$. First row: influence of the Mach number $\Mac$: \#4 (0.4), \#5 (0.8), \#1 (1.1); second row: influence of the Reynolds number $\Rey$: \#2,5 (3300), \#3,6 (8000); third row: influence of a heated environment: \#2 (not heated), \#1 (heated). \lineins: \#1, \linzwei: \#2, \lindrei: \#3, \linvier: \#4, \linfuenf: \#5, \linsechs: \#6, \linb: reference temperatures: wall and ambient respectively total inlet temperature, see table \ref{tab_para}.}
\label{fig:uvr_r}
\end{figure}

The first column of figure \ref{fig:uvr_r} shows the radial velocity $u_r$ at the wall distance of $y/D=0.05$. Starting from the stagnation point it strongly increases due to the stagnation point pressure and reaches a maximum value that is lower then the inlet velocity $u_{\infty}$. In the first row, the influence of the Mach number $\Mac$ is shown. The radial velocity slightly decreases with increasing $\Mac$. Stronger is the influence of the Reynolds number (second row). $u_r$ decreases with increasing $\Rey$. The position of the maximum moves to greater $r/D$. In this plot and for all further comparisons, simulation \#2 and \#3 have to be compared to each other. Simulations \#5 and \#6 are another pair. The pairs (\#2,3) and (\#5,6) differ by the Mach number as well as the ambient and wall temperature (see table \ref{tab_para}). Another noticeable characteristics is that the maximal radial velocity decreases and the position of the maximum moves closer to stagnation point when the environment and the walls are heated (third row of figure \ref{fig:uvr_r}).

In the considered height ($y/D=0.05$) all simulations have a negative axial velocity (around 10 to 14 \% of the inflow velocity) close to the stagnation point (second column of figure \ref{fig:uvr_r}). This means that the flow approaches the wall. Moving in radial direction, this component turns slightly positive (up to 2 percent of inlet velocity). This is a consequence of the thickening of the wall jet: the flow spreads in positive $y$ direction. The radial position ($r/D$) where the direction turns (zero-crossing) decreases with increasing $\Mac$ (first row), decreasing $\Rey$ (second row) and a heated environment (third row). After reaching a maximum value, depending on the configuration, the axial velocity decreases and features weak local maxima. The simulation with the lowest Mach number (\#4) differs from those with higher $\Mac$ in the fact, that the maximum axial velocity is not in stagnation point, but at $r/D\approx 0.32$.

The third and fourth column of figure \ref{fig:uvr_r} show the temperature respectively total temperature profiles of the impinging jets. The temperature profile is influenced by three effects: a) the mixing of the wall jet with the hot environment (cases \#1,4,5,6). This causes the main trend of increasing temperature with increasing $r/D$. b) Heat transfer at the isothermal wall and c) compressibility effects. The first row shows the influence of the Mach number. At low $Ma$ (\#4; $Ma\approx 0.4$), the average temperature $\ol{T}$ and total temperature $\ol{T_t}$ reach the value of the total inlet temperature $T_{t,i}$. At higher Mach number (sub- and supersonic jets) both temperatures in the stagnation point are higher then $T_{t,i}$. This is an indication, that the stagnation point is less instationary when the Mach number is low. An instationary stagnation point means mixing with the surrounding fluid and in case of a present temperature difference, an increasing temperature in the stagnation point.

The effect of compressibility can be seen in the areas with a high (radial) velocity. A high velocity leads to a decreasing temperature, e.g. at $r/D \approx 0.8$, the supersonic case (\#1) features a global minimum at the point where the axial velocity is maximal (first row). The same effect is present for the high subsonic jet (\#5), but of cause weaker. The position of the decreased temperature variates (with respect to $Ma$, $Re$, and the hot environment) in accordance with the position of the high axial velocity.
	
Looking at the total temperature, we see a less strong increase respectively even decreasing temperature at $r/D\approx 1.5$. This is no effect of compressibility (since we look at the total temperature). In the supersonic cases with no hot environment, the temperature even falls below the total inlet temperature. This can happen only due to a heat flux at the isothermal impinging plate. 

\subsection{Wall pressure}
\label{sec:wall_press}

\begin{figure}
\captionsetup[subfigure]{labelformat=empty}
\centering
\subfloat[]{\includegraphics[width=0.245\textwidth]{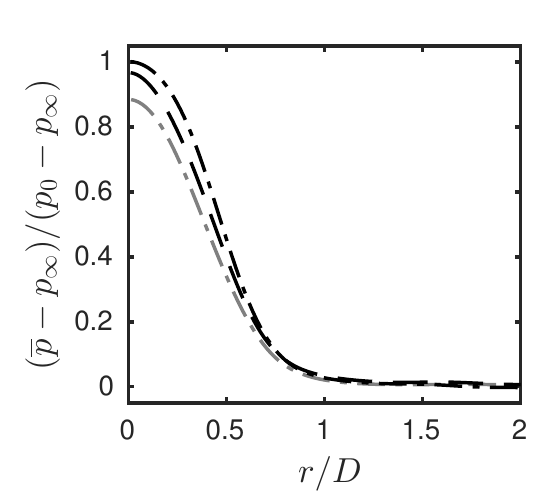}}
\subfloat[]{\includegraphics[width=0.245\textwidth]{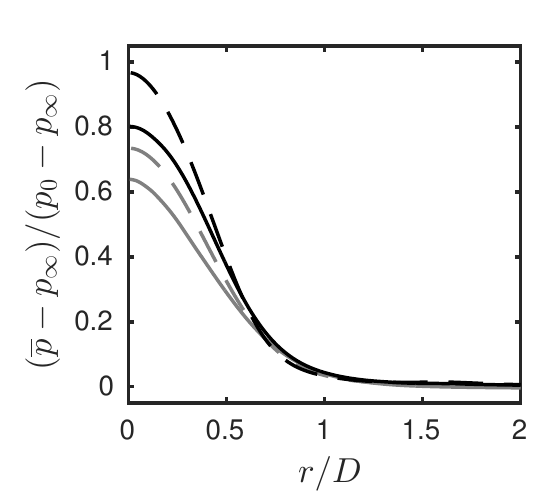}}
\subfloat[]{\includegraphics[width=0.245\textwidth]{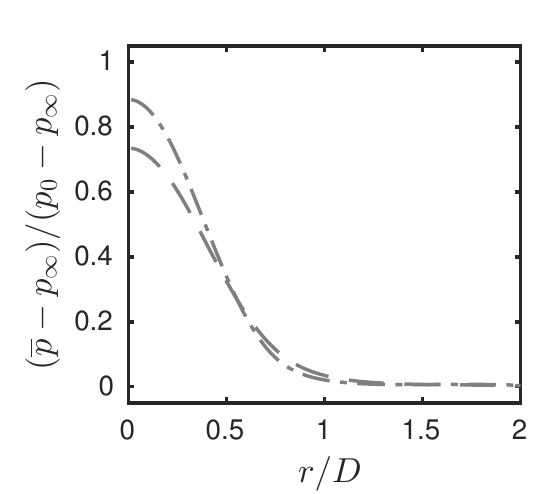}}\\[-8mm]

\subfloat[]{\includegraphics[width=0.245\textwidth]{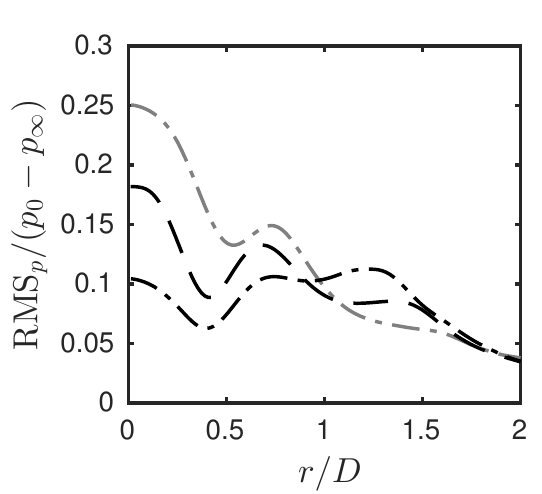}}
\subfloat[]{\includegraphics[width=0.245\textwidth]{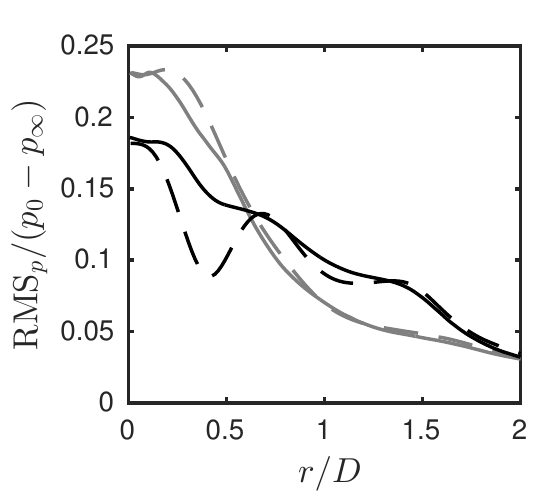}}
\subfloat[]{\includegraphics[width=0.245\textwidth]{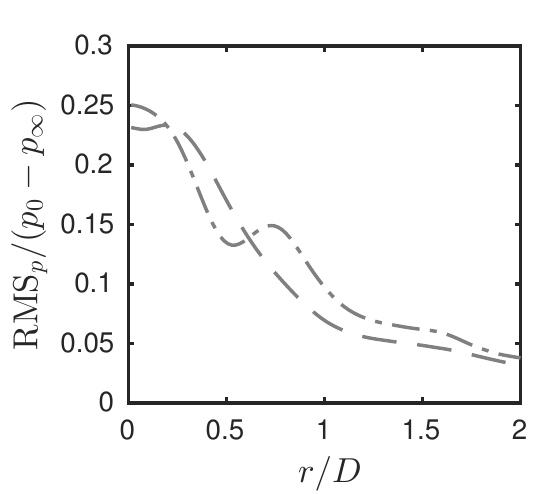}}\\[-6mm]

\caption{Dimensionless pressure at the impinging plate. Left: influence of the Mach number $\Mac$: \#4 (0.4), \#5 (0.8), \#1 (1.1); middle: influence of the Reynolds number $\Rey$: \#2,5 (3300), \#3,6 (8000); right: influence of a heated environment: \#2 (not heated), \#1 (heated). \lineins: \#1, \linzwei: \#2, \lindrei: \#3, \linvier: \#4, \linfuenf: \#5, \linsechs: \#6.}
\label{fig:wall_p}
\end{figure}

In the previous section, it was indicated by the temperature that an impinging jet with a low Mach number has a more stationary stagnation point. This is confirmed by the wall pressure and its RMS at the impinging plate, which is shown in figure \ref{fig:wall_p}. At the low Reynolds number (3300), the pressure recovery is around 1 for the low Mach number ($Ma=0.4$, \#4) and 0.97 at the higher Mach number ($Ma=0.8$, \#5). The supersonic simulation (\#1) reaches a pressure recovery of around 0.9. 

We can make this clear with a simple gedankenexperiment. We approximate the high pressure around the stagnation point as a half unit circle with the flat side to the bottom and add a normally distributed random movement of the center. In case of a very small movement, e.g. with a variance of 0.1, the shape of the function is still approximately a half unit circle and the maximum value is still close to one (0.995). This case represents simulation \#4. If we increase the variance of the random displacement (e.g. \#5) to 0.3, the shape changes and the maximum value decreases to 0.951. Consequently, the two curves have to cross each other at some point. This crossing can be observed between simulations \#4 and \#5 as well as (\#2, \#3) (\#5, \#6) and (\#1, \#2). However no crossing between (\#1, \#4) respectively (\#1, \#5) implies that the lower pressure recovery in the supersonic case is mainly caused by losses due to the shocks present in the free jet region as well as the standoff shocks.

The corresponding raised RMS values indicate an increased movement of the stagnation point. The plots in the middle show that also an increasing Reynolds number leads to larger fluctuations of the stagnation point and consequently to a lower pressure recovery. In contrary, a temperature difference stabilises the stagnation point (right plot). This can be attributed to different modes, as described in \cite{WilkeSesterhenn2016b}.

\subsection{Boundary layer}

The boundary layer is resolved in all cases. For simulations with heat transfer, the maximal $y^+$ value in the first grid point next to the wall is around $0.6$. In order to save computing time, we slightly increased this value (see table \ref{tab_para}) for supersonic impinging jets, since more time steps for the analysis of the acoustics were needed. 

\begin{figure}
\captionsetup[subfigure]{labelformat=empty}
\centering
\subfloat[]{\includegraphics[width=0.245\textwidth]{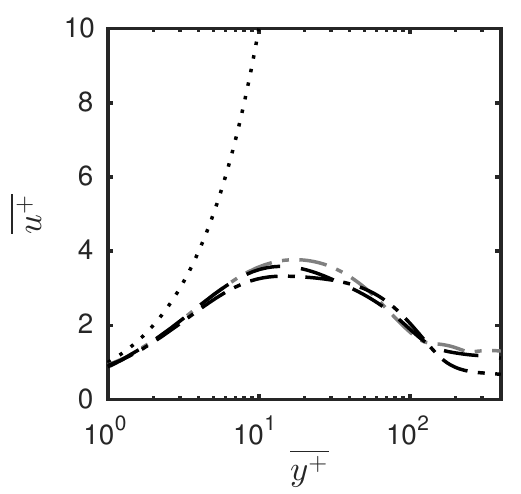}}
\subfloat[]{\includegraphics[width=0.245\textwidth]{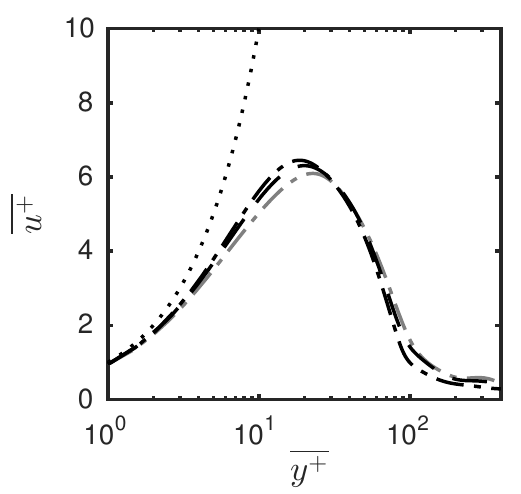}}
\subfloat[]{\includegraphics[width=0.245\textwidth]{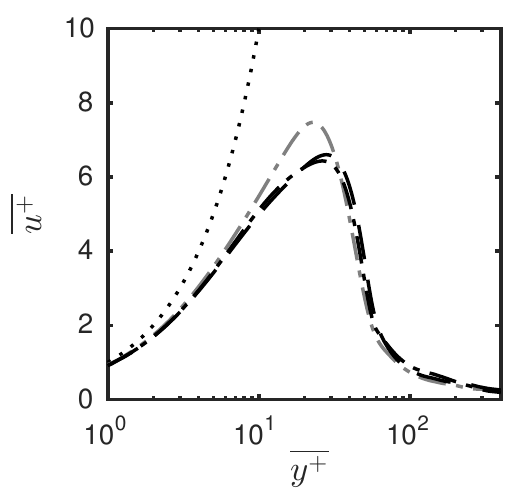}}
\subfloat[]{\includegraphics[width=0.245\textwidth]{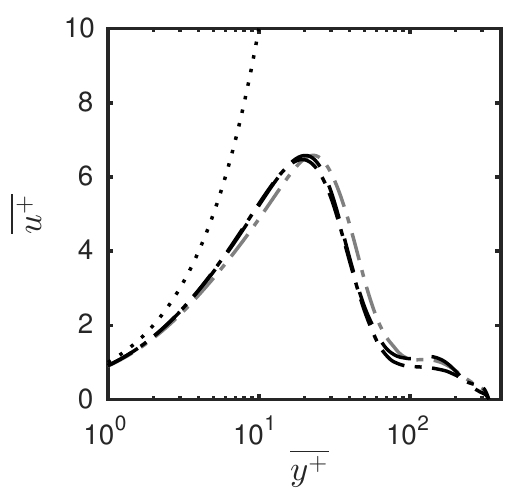}}\\[-8mm]

\subfloat[]{\includegraphics[width=0.245\textwidth]{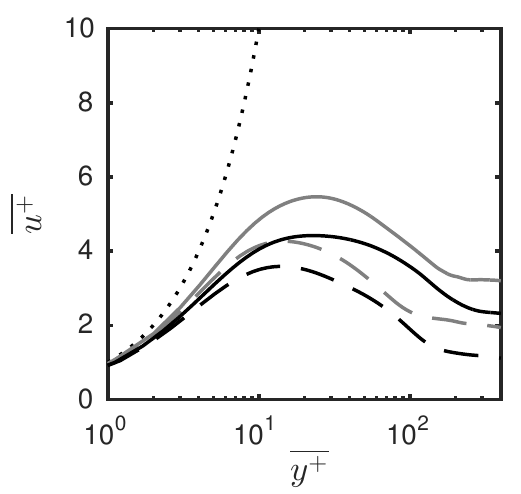}}
\subfloat[]{\includegraphics[width=0.245\textwidth]{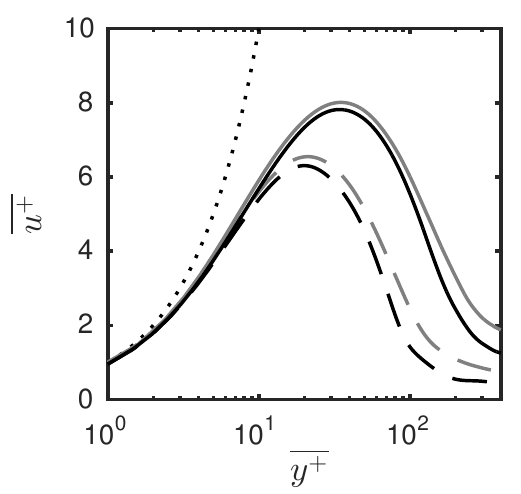}}
\subfloat[]{\includegraphics[width=0.245\textwidth]{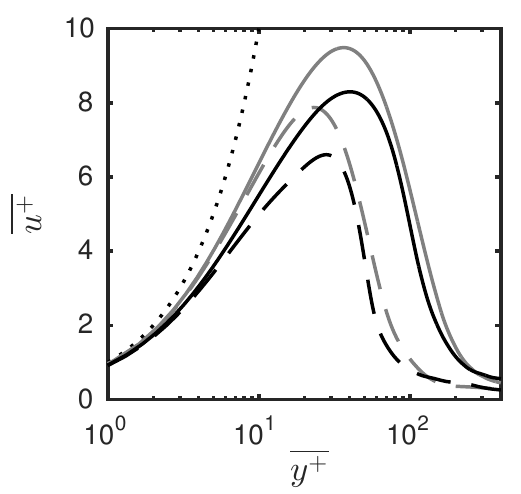}}
\subfloat[]{\includegraphics[width=0.245\textwidth]{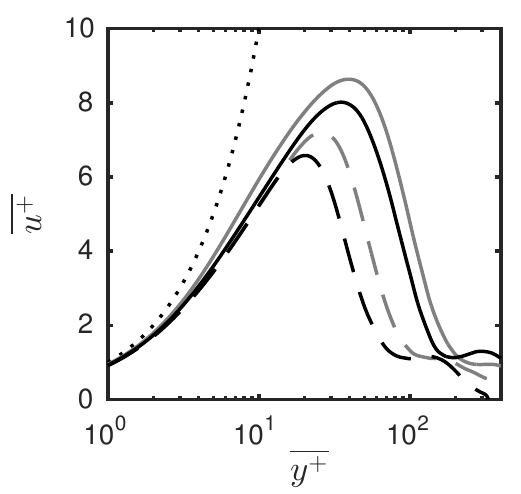}}\\[-8mm]

\subfloat[]{\includegraphics[width=0.245\textwidth]{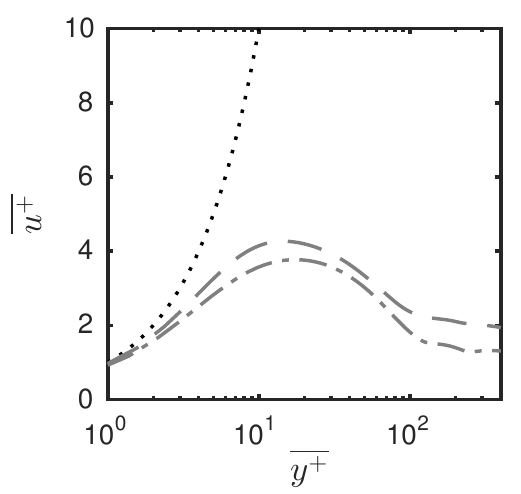}}
\subfloat[]{\includegraphics[width=0.245\textwidth]{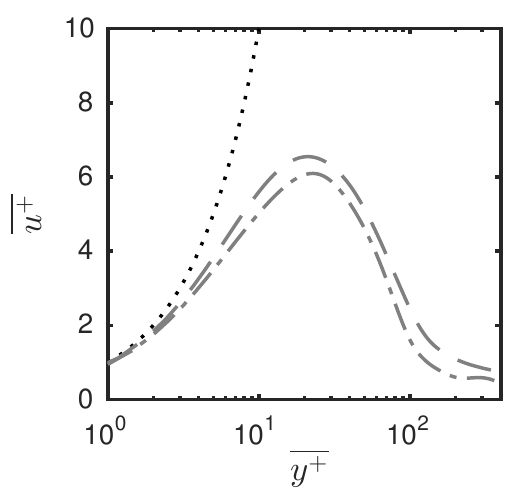}}
\subfloat[]{\includegraphics[width=0.245\textwidth]{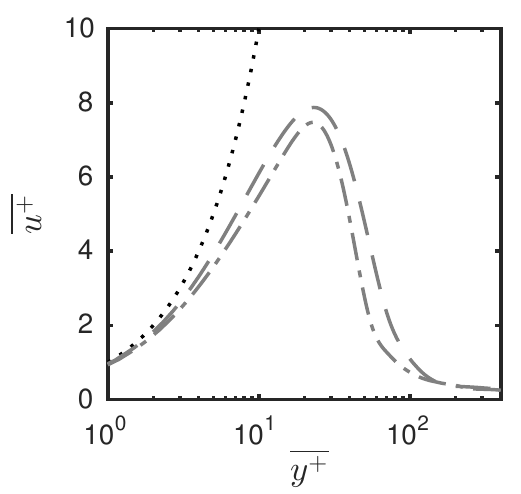}}
\subfloat[]{\includegraphics[width=0.245\textwidth]{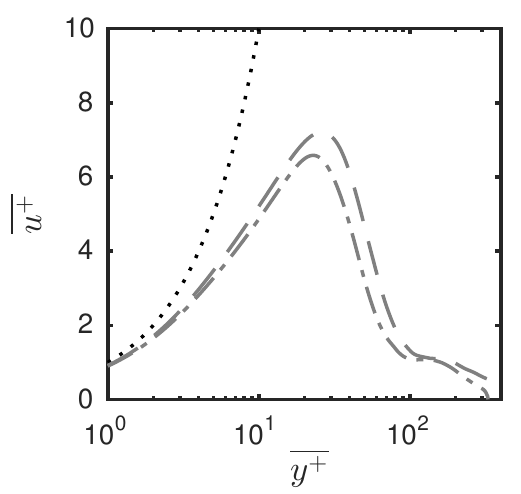}}\\[-6mm]
\caption{Velocity boundary layer for different radial positions (left to right: $r/D=0.3$, $0.8$, $1.4$ and $3.5$). First row: influence of the Mach number $\Mac$: \#4 (0.4), \#5 (0.8), \#1 (1.1); second row: influence of the Reynolds number $\Rey$: \#2,5 (3300), \#3,6 (8000); third row: influence of a heated environment: \#2 (not heated), \#1 (heated). \lineins: \#1, \linzwei: \#2, \lindrei: \#3, \linvier: \#4, \linfuenf: \#5, \linsechs: \#6, \linpk: $u^+=y^+$.}
\label{fig:BL_velo}
\end{figure}

Figure \ref{fig:BL_velo} shows the velocity boundary layer of the impinging jets. $y^+$ and $u^+$ are the dimensionless wall distance and velocity. The $u^+$-profile of the wall jet is for all computations and $r/D$ lower then the solution of channel flow. The maximum is caused by the fact, that the wall jet has a finite thickness. The fluid above the wall jet is almost at rest, neglecting a slight recirculation. The profile is strongest influenced by the Reynolds number, followed by the heated environment. The Mach number has a small influence, except for $r/D=1.4$. Increasing $\Mac$ leads to a movement of the $u^+$-maximum to higher values of $y^+$. Also here $r/D=1.4$ forms an exception. For radial distance other than $r/D=0.8$, the dimensionless velocity increases with increasing Mach number. The entire profile is raised with the Reynolds number. A heated environment has the impact that $u^+$ decreases in the entire domain. The vertical position ($y^+$) of the $u^+$-maximum increases for $r/D=0.3$ and $0.8$ and then decreases for larger radial distances. If we compare the profiles regarding the radial distance, the maximal values of $y^+$ and $u^+$ increase until $r/D=1.4$ and then decrease slightly. Additionally, the drop after the maximum gets sharper until $r/D=1.4$.

Contrary to the velocity, the thermal boundary layer profile can be below or exceed the channel flow profile ($T^+=\Pran \; y^+$). In the case of the heated environment, the curves are close to $T^+=\Pran \; y^+$, depending on the radial position either until $y^+\approx 5$ or $y^+\approx 10$. The influence of the Mach number is much stronger than in velocity profile. The plots in the first column at $r/D=0.3$ are in the mixing layer. That's why $T^+$ increases again at high values of $y^+$. Here, we concentrate on the range before ($y^+ \lesssim 70$). Increasing Mach numbers lead to increasing values of the dimensionless temperature until the maximum is reached, except for the radial distance $r/D=1.4$. No trend regarding the position of the maximum can be determined. An increasing Reynolds number leads to increasing values of $T^+$ and the radial position, where the maximum is observed. However, when the environment is not heated, the Reynolds number has almost no effect until the maximum is reached and until $r/D \leq 0.8$. The heated environment leads to much higher values of $T^+$ in the entire domain.

\begin{figure}
\captionsetup[subfigure]{labelformat=empty}
\centering
\subfloat[]{\includegraphics[width=0.245\textwidth]{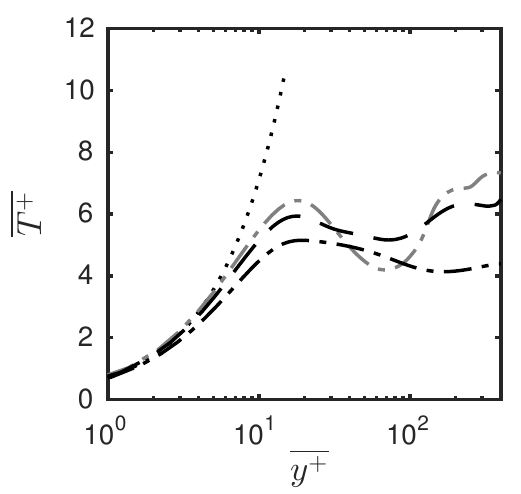}}
\subfloat[]{\includegraphics[width=0.245\textwidth]{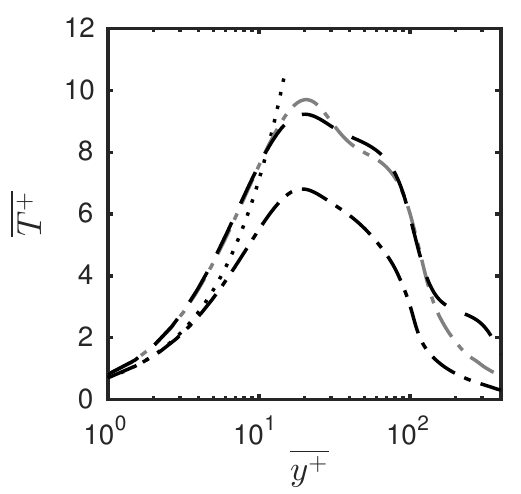}}
\subfloat[]{\includegraphics[width=0.245\textwidth]{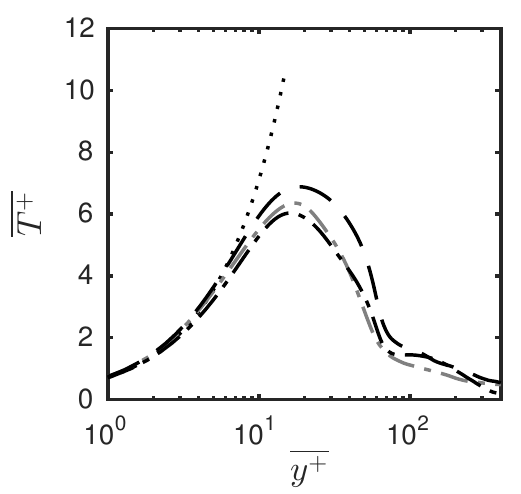}}
\subfloat[]{\includegraphics[width=0.245\textwidth]{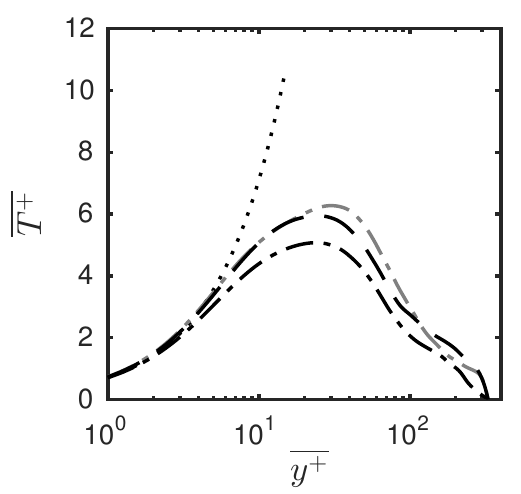}}\\[-8mm]

\subfloat[]{\includegraphics[width=0.245\textwidth]{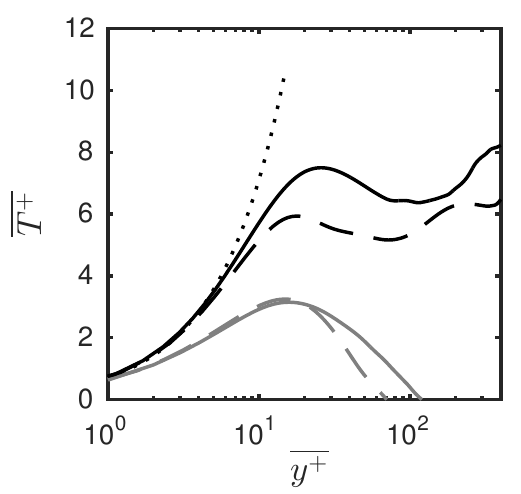}}
\subfloat[]{\includegraphics[width=0.245\textwidth]{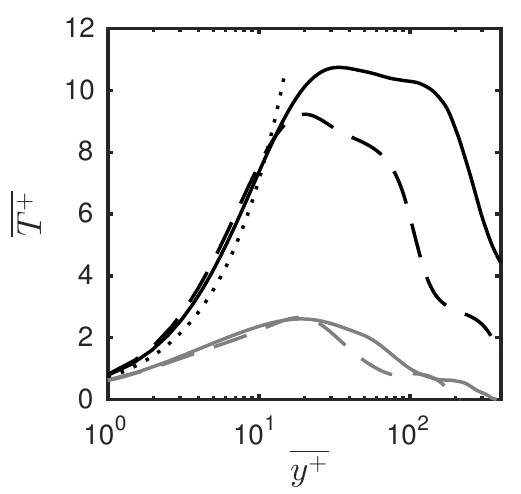}}
\subfloat[]{\includegraphics[width=0.245\textwidth]{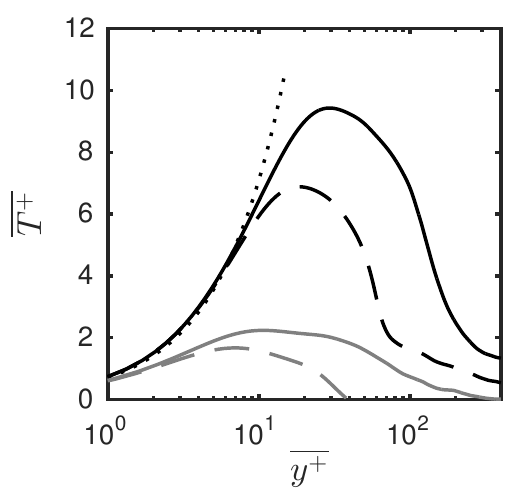}}
\subfloat[]{\includegraphics[width=0.245\textwidth]{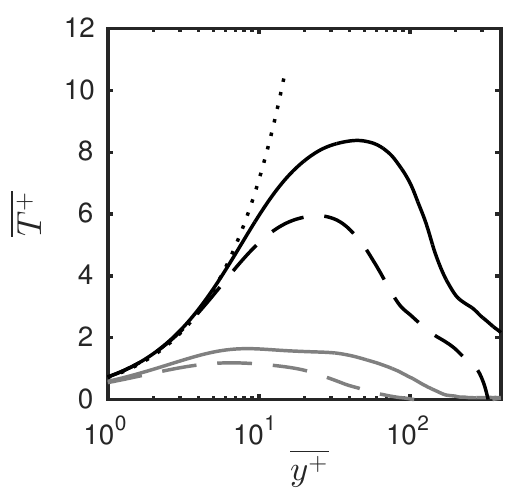}}\\[-8mm]

\subfloat[]{\includegraphics[width=0.245\textwidth]{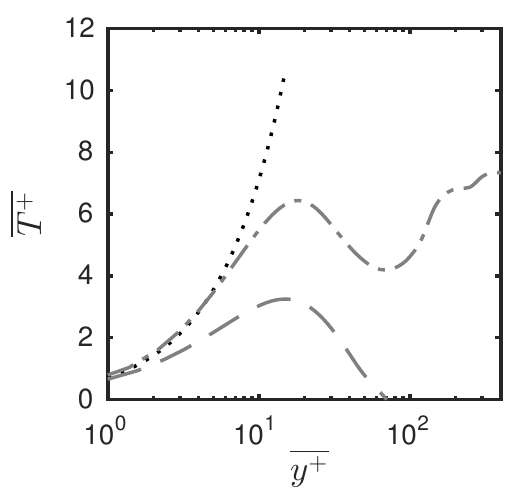}}
\subfloat[]{\includegraphics[width=0.245\textwidth]{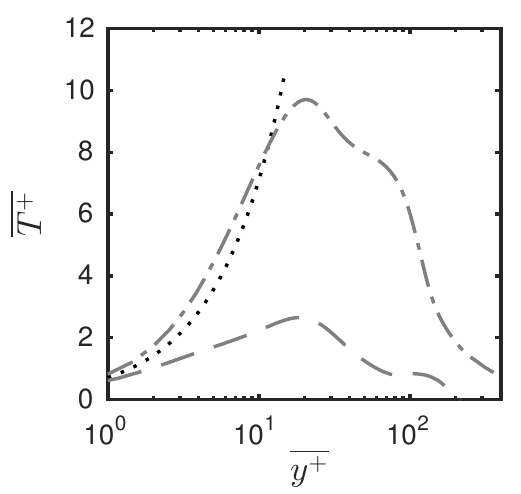}}
\subfloat[]{\includegraphics[width=0.245\textwidth]{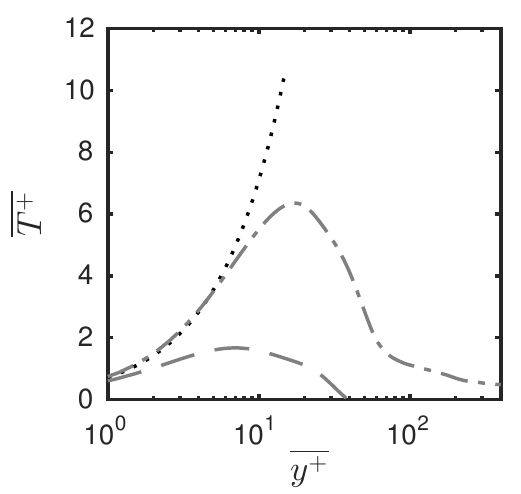}}
\subfloat[]{\includegraphics[width=0.245\textwidth]{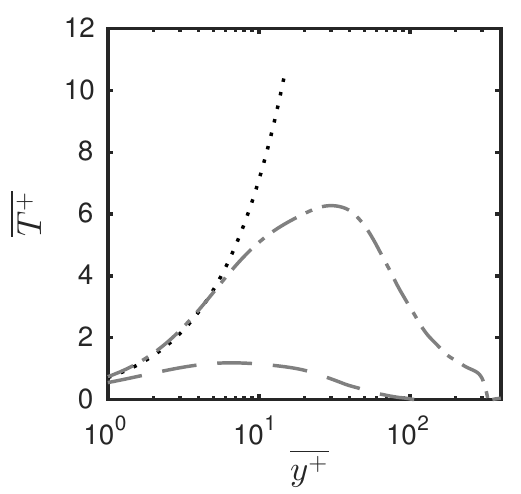}}\\[-6mm]
\caption{Temperature boundary layer for different radial positions (left to right: $r/D=0.3$, $0.8$, $1.4$ and $3.5$). First row: influence of the Mach number $\Mac$: \#4 (0.4), \#5 (0.8), \#1 (1.1); second row: influence of the Reynolds number $\Rey$: \#2,5 (3300), \#3,6 (8000); third row: influence of a heated environment: \#2 (not heated), \#1 (heated). \lineins: \#1, \linzwei: \#2, \lindrei: \#3, \linvier: \#4, \linfuenf: \#5, \linsechs: \#6, \linpk: $T^+=\Pran \; y^+$.}
\label{fig:BL_temp}
\end{figure}


\section{Reynolds analogies and correlations}

Since impinging jets are used among other things for the cooling of hot surfaces, an accurate computation of the heat transfer is needed. Furthermore, relations between quantities affiliated to heat transfer and quantities affiliated to momentum transfer are of great interest.

\subsection{Mean field}
\label{sec:RA_mean}

The heat transfer at the impinging plate is quantified by the Nusselt number:
\begin{equation}
	\Nus= \frac{D}{\Delta T} \cdot \left. \pd{T}{y} \right|_W  = \dot{q}_W \frac{D \Pran}{\Delta T c_p \eta} \qquad .
\end{equation}

\noindent $D$ is the inlet diameter, $\Delta T$ is the difference between the total inlet temperature $T_{t,i}$ and the wall temperature $T_W$. $\Pran, c_p, \eta$ and $q_W$ are the Prandtl number, the ratio of specific heats, the dynamic viscosity and the heat flux in wall normal direction at the wall. Reynolds discovered that the similarity of the momentum and energy equation for incompressible laminar boundary layers can be used to approximate the heat transfer with the use of the fluid friction, see \cite{Kakag1995}:
\begin{equation}
	\Sta = \frac{\Nus}{\Rey \Pran} \approx = \frac{C_f}{2} = \frac{\tau_W}{\rho u_{\infty}^2}\qquad .
	\label{eq:RA}
\end{equation}

\begin{figure}
\centering
\subfloat[Influence of $\Mac$: \newline \#4 (0.4), \#5 (0.8), \#1 (1.1)]{\includegraphics[width=0.32\textwidth]{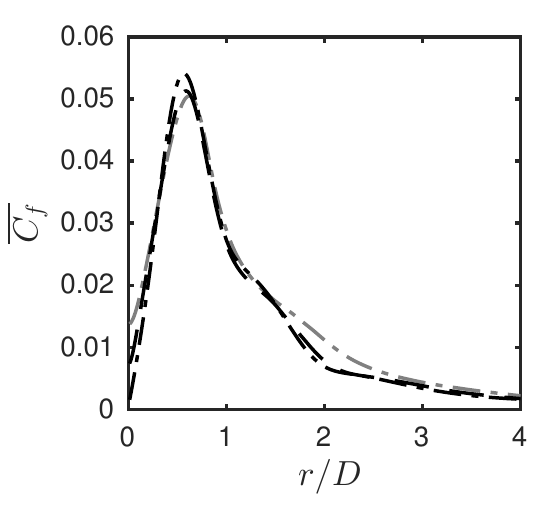}}
\subfloat[Influence of $\Rey$: \newline \#2,5 (3300), \#3,6 (8000)]{\includegraphics[width=0.32\textwidth]{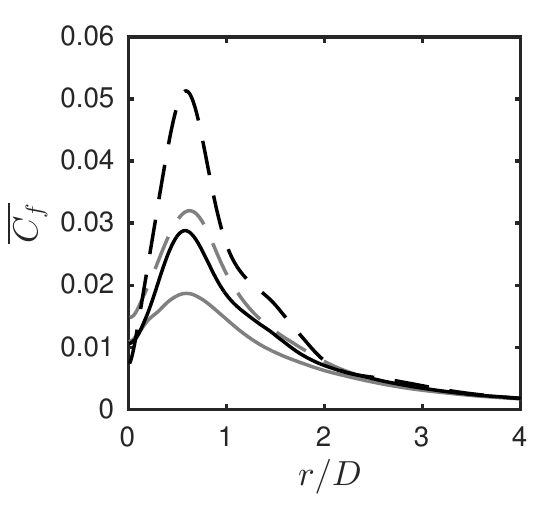}}
\subfloat[Influence of ambient temperature: \newline \#2 (not heated), \#1 (heated)]{\includegraphics[width=0.32\textwidth]{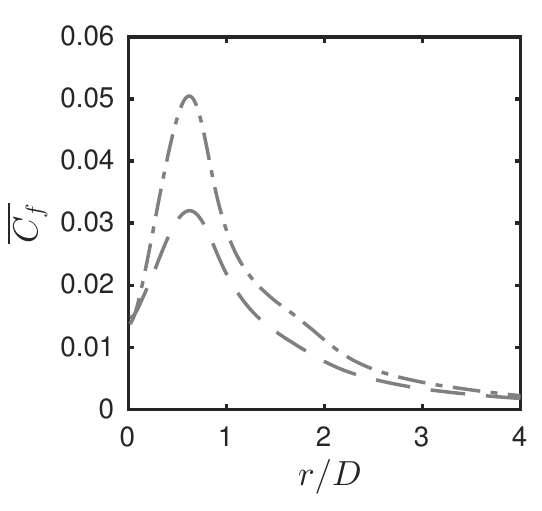}}

\subfloat[Influence of $\Mac$: \newline \#4 (0.4), \#5 (0.8), \#1 (1.1)]{\includegraphics[width=0.32\textwidth]{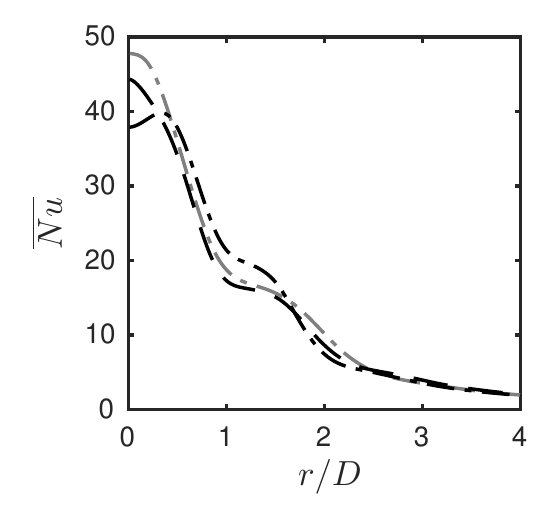}}
\subfloat[Influence of $\Rey$: \newline \#5 (3300), \#6 (8000)]{\includegraphics[width=0.32\textwidth]{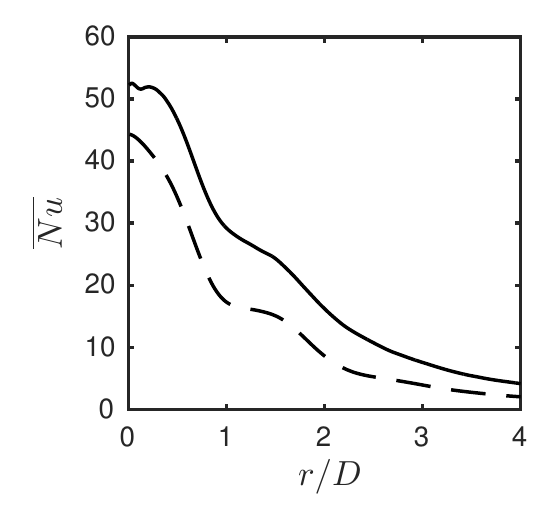}}
\subfloat[Scaled $\Rey$: \newline \#5 (3300), \#6 (8000)]{\includegraphics[width=0.32\textwidth]{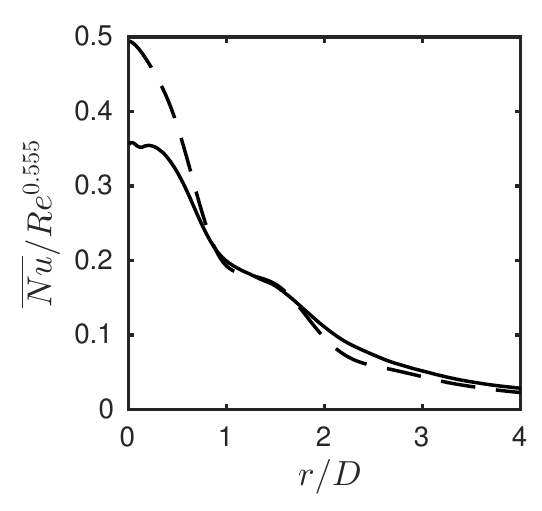}}
\caption{Skin friction factor $C_f$ (a-c) and Nusselt number $\Nus$ (d-f), \lineins: \#1, \linzwei: \#2, \lindrei: \#3, \linvier: \#4, \linfuenf: \#5, \linsechs: \#6.}
\label{fig:Cf_Nu}
\end{figure}

\noindent This is the well known Reynolds analogy. The assumption made is a Prandtl number equal to one. $C_f$ and $\tau_W$ are the skin friction factor and the wall shear stress. This equation was modified by \cite{ChiltonColburn1934} based on experimental data:

\begin{equation}
	  \Sta \Pran^{2/3} = \frac{\Nus}{\Rey \Pran^{1/3}}\approx \frac{C_f}{2}
	  \label{eq:CC}
\end{equation}

\noindent and considers Prandtl numbers different from one. Equation \ref{eq:CC} is referred to as the Chilton Colburn analogy. The first row of figure \ref{fig:Cf_Nu} shows the skin friction coefficients of the conducted simulations. The left plot indicates that $C_f$ is almost independent of the Mach number in the range of $0.4 \leq \Mac \leq 1.1$. As expected, an increasing Reynolds number leads to a decreasing skin friction factor. However the shape of the profile is not affected. In the middle plot, two pairs of simulations are shown: subsonic impinging jets (black solid line and black dashed line) and supersonic cases (grey solid line and grey dashed line). The difference between the pairs can be explained using the right plot. Comparing a jet with equal total inlet temperature $T_{t,i}$, ambient $T_{\infty}$ and wall temperature $T_W$ to another one with $T_{\infty} = T_W > T_{t,i}$, it can be seen that the skin friction factor increases due to the heated environment. The total inlet temperature has been kept constant.

In the second row of figure \ref{fig:Cf_Nu}, the Nusselt number profiles of the simulations with $T_{\infty} = T_W > T_{t,i}$ are compared (otherwise $\Delta T =0$). The influence of the Mach number (left) is strongest in the vicinity of the stagnation point. The simulation with the low compressibility ($\Mac \approx$ 0.4) has the global maximum at around $r/D=0.3$ and not at the axis. $\Nus$ increases with increasing $\Mac$ at the stagnation point. This is an indication of stronger fluctuations and an increased contribution of the turbulent heat flux. This observation is consistent with the behaviour of the pressure at the impinging plate, as described in section \ref{sec:wall_press}. All simulations feature a secondary local maximum respectively shoulder or saturation zone around $r/D=1.5$. Well known are Nusselt number correlations of the shape $\Nus \sim \Rey^n$. Also the Prandtl number and geometrical parameters can be included. According to \cite{LeeLee1999} the exponents for plate distances $h/D$ of 4 respectively 6 are: $n= 0.53$; $n= 0.58$. For the presently investigated simulations of $h/D=5$ the exponent of $n= 0.555$ was chosen. The influence of the Reynolds number is illustrated in the second row of figure \ref{fig:Cf_Nu} (middle and right). As expected, the heat transfer increases with $\Rey$. The scaling fits away from the stagnation point $r/D \gtrsim 1$. According to the simple correlation, the heat transfer in the stagnation point area of the higher Reynolds number (8000) is weaker than in the case of $Re=3300$.

\begin{figure}
\captionsetup[subfigure]{labelformat=empty}
\centering
\subfloat[]{\includegraphics[width=0.32\textwidth]{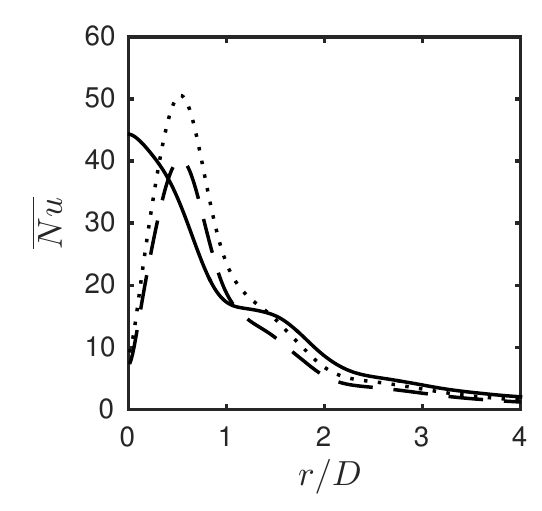}}
\subfloat[]{\includegraphics[width=0.32\textwidth]{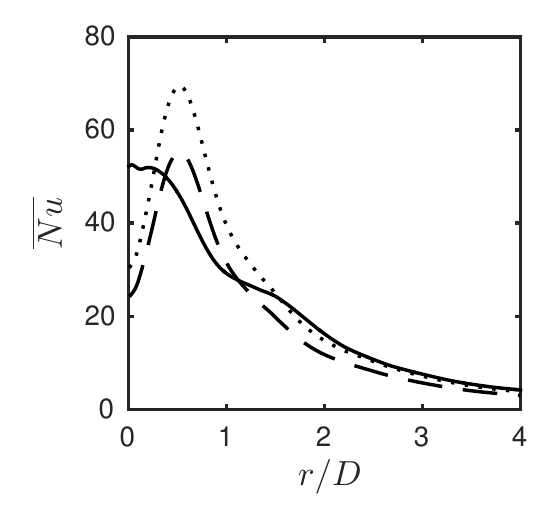}}\\[-6mm]
\caption{Heat transfer at the impinging plate (Nusselt number). Left: simulation \#5, right: \#6, \link: DNS data, \lindk: Reynolds analogy (RA), \linpk: Chilton Colburn analogy (CCA).}
\label{fig:RA_CC}
\end{figure}

The Reynolds (RA) and the Chilton Colburn analogies (CCA) are shown in figure \ref{fig:RA_CC}. As they are developed for wall-bounded flows, the analogies are not suitable for the stagnation point region. In this region the relative error is between around -80\% and -40\%. At the position where $C_f$ reaches its maximum, both analogies overpredict the heat transfer: $\approx +30\%$ (RA) respectively $\approx +65\%$ (CCA). Farther away from the axis $r/D \gtrsim 2$ the analogies fit much better. The best agreement is found in the case of the higher Reynolds number using the CCA. Here the error is between -9\% and -4\%.

Other relations involving the mean temperature and the mean velocity were developed by \cite{Crocco1932} and \cite{Busemann1931}, \cite{Walz1962} and \cite{ZhangBi2014}. In \cite{ZhangBi2014} the derivation is explained in detail. All relations have the common form:
\begin{equation}
	\frac{\ol{T}}{\ol{T}_{\delta}} = \frac{\ol{T}_W}{\ol{T}_{\delta}} + \frac{T_r - \ol{T}_W}{\ol{T}_{\delta}} \; \frac{\ol{u_r}}{\ol{u_r}_{\delta}} + \frac{\ol{T}_{\delta} - T_r}{\ol{T}_{\delta}} \rk{\frac{\ol{u_r}}{\ol{u_r}_{\delta}}}^2 \qquad ,
\end{equation}
 
\noindent with
\begin{equation}
	T_r = \ol{T}_{\delta} + r \frac{\ol{u_r}_{\delta}^2}{2 c_p} \qquad .
\end{equation}

\noindent The recovery factor $r$ changes according to the authors. For $r=1$ the Crocco-Busemann relation (CBR) is derived. In the Walz's equation or modified Crocco-Busemann relation $r=0.88$. The generalized Reynolds analogy (GRA) proposed by \cite{ZhangBi2014} uses the general recovery factor according to equation \ref{eq:r_zhang}:
\begin{equation}
	r=\rk{\ol{T}_W-\ol{T}_{\delta}} \frac{2 c_p}{\ol{u_r}_{\delta}^2} - \frac{2 \Pran}{\ol{u_r}_{\delta}} \; \frac{\ol{q}_W}{\ol{\tau_W}} \qquad .
	\label{eq:r_zhang}
\end{equation}

\noindent Those three relations were tested for the impinging jet. The difference between the Crocco-Busemann and Walz's equation was found to be negligible for the present simulations. For the reason of lucidity, the approximation according to Walz is not shown.

\begin{figure}
\captionsetup[subfigure]{labelformat=empty}
\centering
\subfloat[]{\includegraphics[scale=0.8]{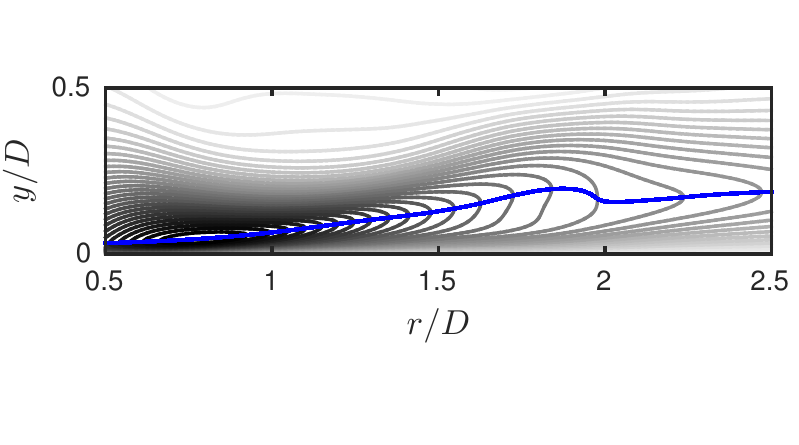}}
\subfloat[]{\includegraphics[scale=0.8]{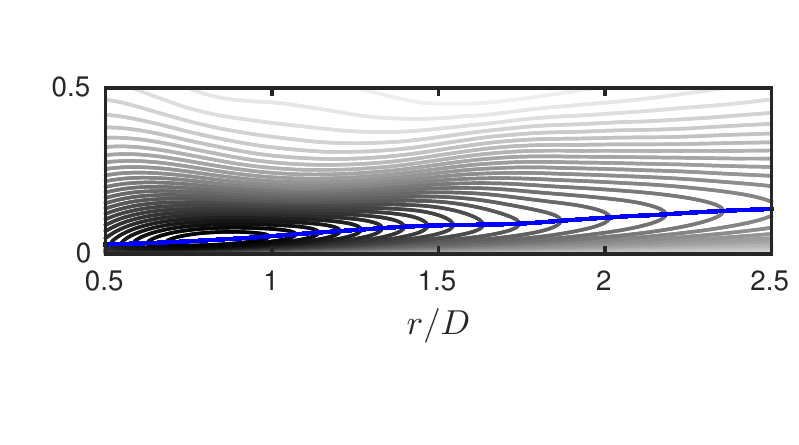}}\\[-20mm]

\subfloat[]{\includegraphics[scale=0.8]{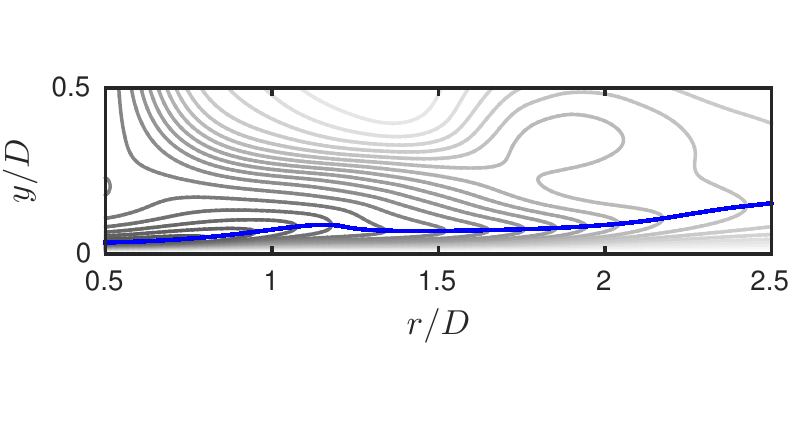}}
\subfloat[]{\includegraphics[scale=0.8]{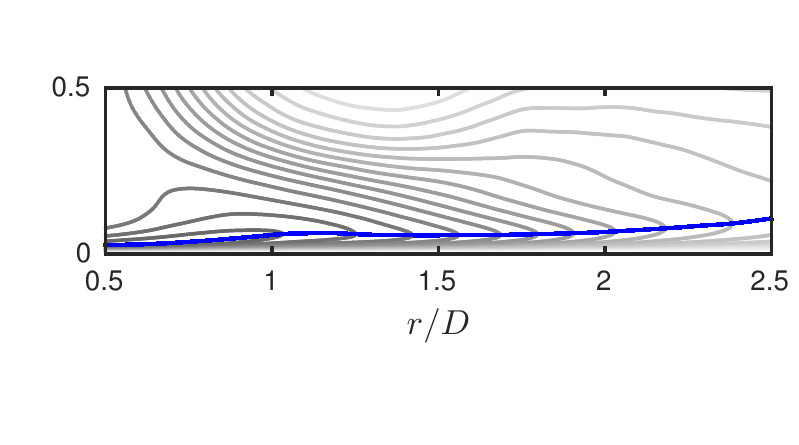}}\\[-8mm]
\caption{Boundary layer thickness (\linb). Top: radial velocity $\ol{u_r}_{\delta}$, bottom: temperature $\ol{T}_{\delta}$, left: simulation \#5, right: \#6. In the background are contour lines of $\ol{u_r}$ respectively $\ol{T}$.}
\label{fig:deltaTur}
\end{figure}

In figure \ref{fig:deltaTur} the boundary layer thickness of the radial (wall parallel) velocity $\ol{u_r}_{\delta}$ and temperature $\ol{T}_{\delta}$ are shown. Both boundary layers increase with increasing radial distance in all cases. Additionally to the thinner boundary layer in the case of the higher Reynolds number \#6 ($\Rey=8000$), compared to \#5 ($\Rey=3300$), another difference occurs. At $\Rey=3300$, the velocity boundary layer features a local decrease at $1.9 \leq r/D \leq 2$. In this area, the average radial velocity is strongly influenced by the periodical movement of primary and secondary vortex rings. The drop in $\ol{u_r}_{\delta}$ is caused by the separation of the vortex pair at this location. At higher Reynolds number, the vortex rings occur with the same frequency, but due to the higher level of turbulence, the radial position of the separation as well as the shape of the vortices vary stronger as at $\Rey=3300$. Consequently, no exact repetitive location of separation is present and no drop in the velocity boundary layer occurs.

\begin{figure}
\captionsetup[subfigure]{labelformat=empty}
\centering
\subfloat[]{\includegraphics[width=0.33\textwidth]{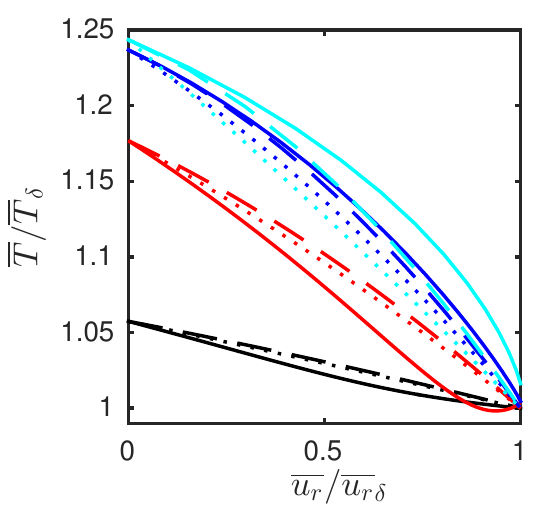}}
\subfloat[]{\includegraphics[width=0.33\textwidth]{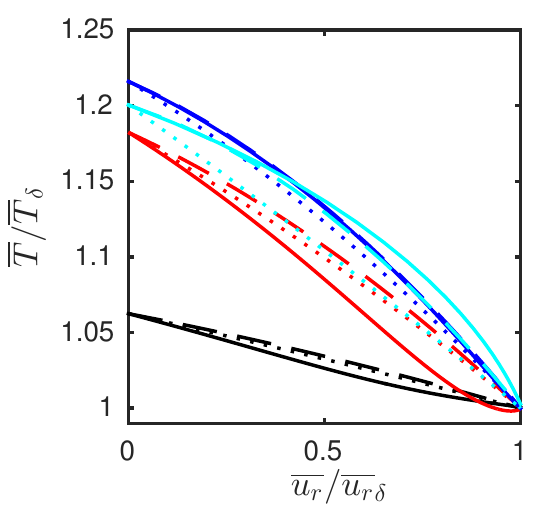}}\\[-6mm]
\caption{Relation between mean temperature and mean velocity in the wall jet. Left: simulation \#5, right: \#6.
colors (online): \linc: $r/D=0.3$, \linb: $r/D=0.8$, \linr: $r/D=1.4$, \link: $r/D=3.5$.
pattern: \link: DNS data, \linpk: Crocco-Busemann relation (CBR) \lindk: generalized Reynolds analogy (GRA).}
\label{fig:GRA_mean}
\end{figure}

Figure \ref{fig:GRA_mean} shows the DNS data compared to the approximations of Crocco-Busemann and the GRA. The mean temperature and mean wall-parallel (radial) velocity are normalised by the values at the edge of the boundary layer (subscript $\delta=\delta_{99}$) as in \cite{ZhangBi2014}. For both Reynolds numbers the GRA fits better than the CBR for radial positions close to the stagnation point ($r/D=0.3$ and $r/D=0.8$). Farther away ($r/D=1.4$ and $r/D=3.5$), the opposite can be observed. This is a consequence of different curvatures of the DNS profiles and the fact that the scaled mean temperature is always predicted higher according to the GRA. Further can be ascertained that for $Re=8000$ and $r/D=0.8$, the GRA gives a precise prediction of the temperature field. At this radial position, the radial velocity has its maximum. Given that no de- and acceleration is present, the conditions are most similar to canonical compressible wall-bounded turbulent flows (CCWTF) for which the relation was developed. In CCWTFs the flow can be approximated as quasi-one-dimensional. Examples for such flows are pipes and channels.

\subsection{Fluctuations}

Additional to the relation between the mean temperature and velocity, \cite{ZhangBi2014} also derived a general analogy for the fluctuations:
\begin{equation}
	T' - \frac{1}{\ol{\Pran_t}} \pd{\ol{T}}{\ol{u}} u' + \phi' - \frac{\ol{\rk{\rho v}'\phi'}}{\ol{\rk{\rho v}' u'}} u' =0
	\label{eq:Tstust}
\end{equation}

\noindent where $\phi'$ is a residual temperature that need to be modelled. The proposed model that was chosen by \cite{ZhangBi2014} for ``convenience'' is:
\begin{equation}
	\phi' = \frac{\ol{\rk{\rho v}'\phi'}}{\ol{\rk{\rho v}' u'}} u'
\end{equation}

\noindent so that eq. \ref{eq:Tstust} is reduced to:
\begin{equation}
	T' = \frac{1}{\ol{\Pran_t}} \pd{\ol{T}}{\ol{u}} u' \qquad .
	\label{eq:GRA_Tst}
\end{equation}

\noindent \cite{ZhangBi2014} describe further that equation \ref{eq:GRA_Tst} is not valid, but the RMS of $T'$ and $u'$ can be approximated in a similar way:
\begin{equation}
	\sqrt{\ol{T'^2}} \approx \left| \frac{1}{\ol{\Pran_t}}\pd{\ol{T}}{\ol{u}} \right|  \sqrt{\ol{u'^2}} \qquad \text{or} \qquad \sqrt{\ol{T'^2}} \approx \pm \frac{\ol{\rk{\rho v}' T'}}{\ol{\rk{\rho v}' u'}} \sqrt{\ol{u'^2}} \qquad .
	\label{eq:GRA_fluctallg}
\end{equation}

\begin{figure}
\captionsetup[subfigure]{labelformat=empty}
\centering
\subfloat[]{\includegraphics[width=0.245\textwidth]{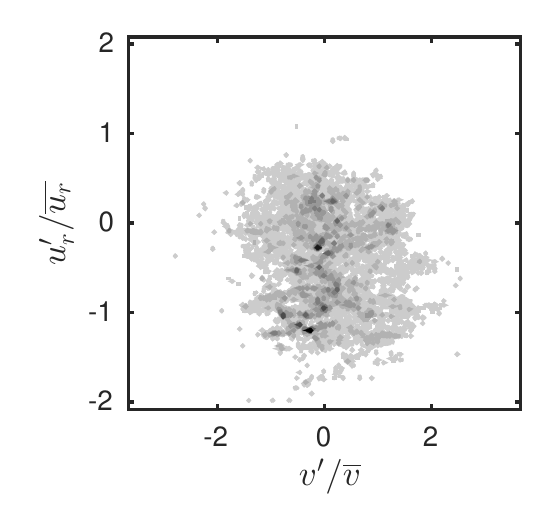}}
\subfloat[]{\includegraphics[width=0.245\textwidth]{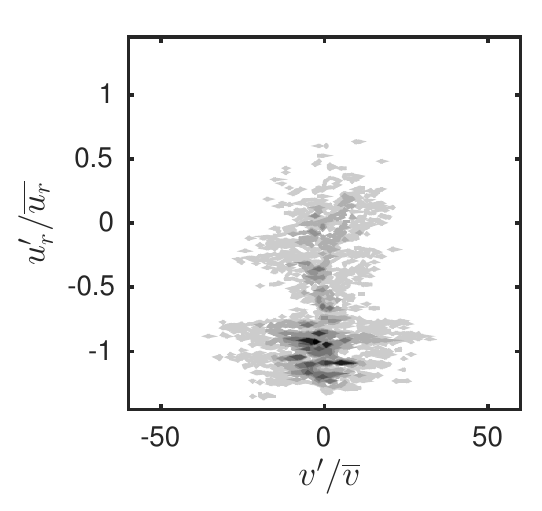}}
\subfloat[]{\includegraphics[width=0.245\textwidth]{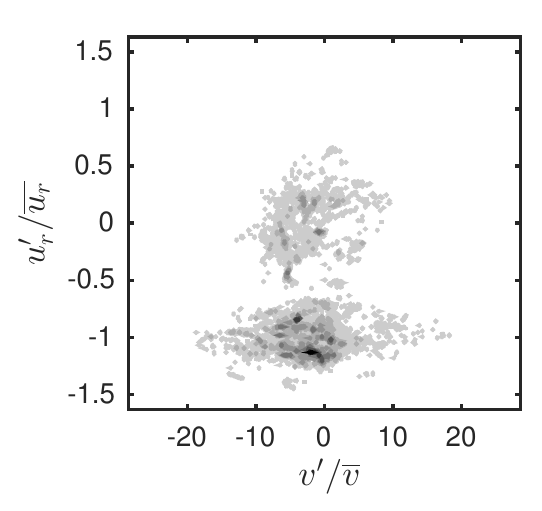}}
\subfloat[]{\includegraphics[width=0.245\textwidth]{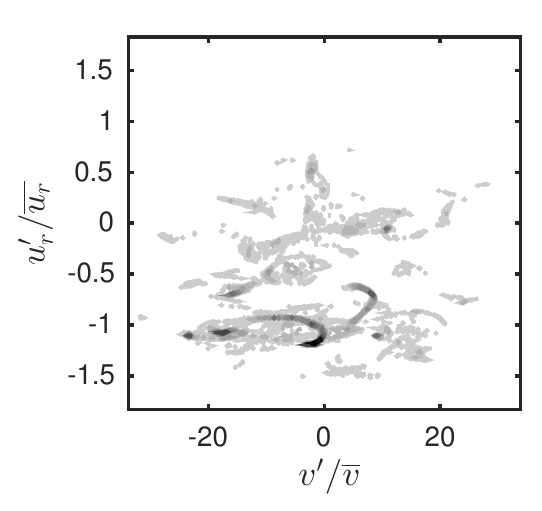}}\\[-8mm]

\subfloat[]{\includegraphics[width=0.245\textwidth]{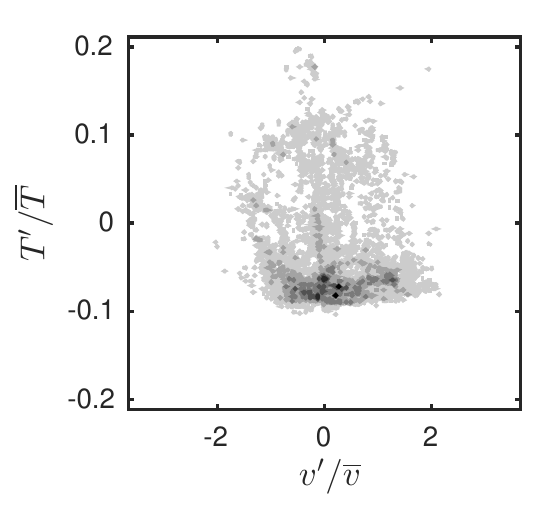}}
\subfloat[]{\includegraphics[width=0.245\textwidth]{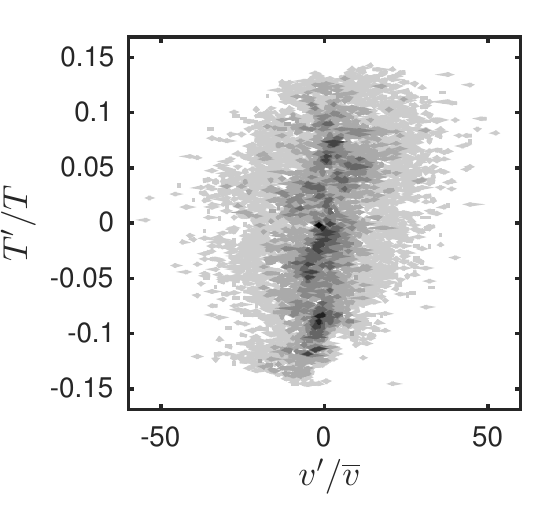}}
\subfloat[]{\includegraphics[width=0.245\textwidth]{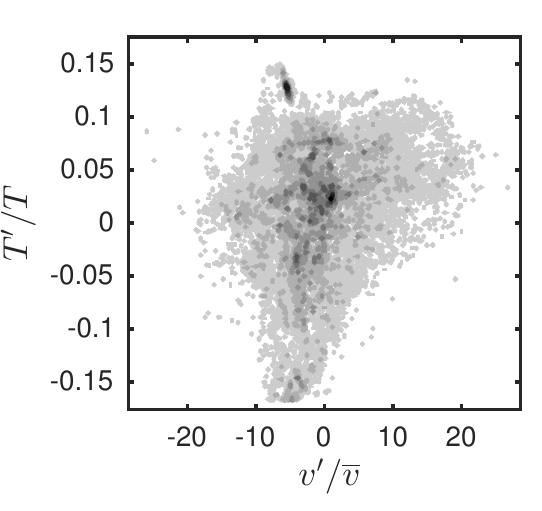}}
\subfloat[]{\includegraphics[width=0.245\textwidth]{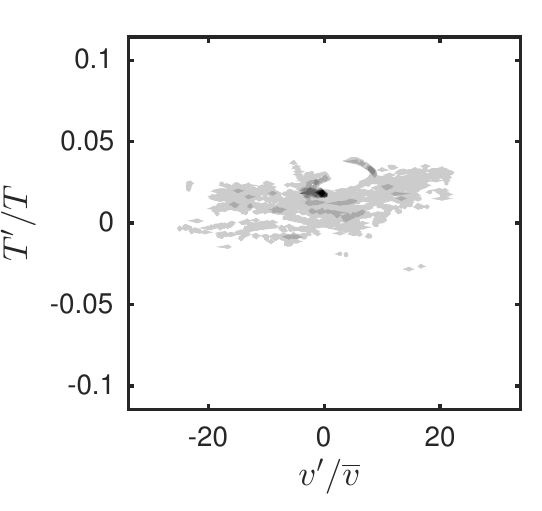}}\\[-8mm]

\caption{Scatter plots of radial velocity (first row) and temperature (second row) fluctuation against the axial velocity fluctuation at $13 \leq y^+ \leq 17$ for different radial positions (left to right: $r/D=0.3$, $0.8$, $1.4$ and $3.5$), simulation \#6.}
\label{fig:scat_vurT6_yp15}
\end{figure}

\noindent The plus sign applies to the flow region where the wall-normal gradients of the mean temperature and velocity have the same sign. The minus sign applies to the opposite situation. This approximation fails in the case of the impinging jet. In the boundary layer the term $\ol{\rk{\rho v}' u'}$ changes its sign. The approximation delivers huge values of $\sqrt{\ol{T'^2}}$ in the vicinity of the zero-crossing. A further approximation is suggested by \cite{ZhangBi2014} that reduces the connection between temperature and velocity fluctuations to $R_{v' u'} \approx R_{v' T'}$. Where R is the correlation coefficient. Also this approximation is invalid for the impinging jet. Figure \ref{fig:scat_vurT6_yp15} exemplary shows scatter plots of simulation \#6 ($\Rey=8000, \Mac \approx 0.8$) in the boundary layer at $13 \leq y^+ \leq 17$. In order to improve the rendering, the data was classified into 250 segments for each variable covering 95\% of the velocity and 99\% of all other fluctuations. The database consists of 550 equally spaced snapshots (symmetry planes) out of 175000 that have been computed after the flow reached its quasi-stationary state. The corresponding correlation coefficients are given in table \ref{tab:corrkoeff} for the same $y^+$ value and additional for $38 \leq y^+ \leq 42$. It can be seen that neither the radial velocity nor the temperature is correlated to the axial velocity. Despite both coefficients are close to zero, it cannot be said that they are approximatively equal.

\subsection{Additional correlations}

The derivation of Reynolds analogies is closely related to the development of models for compressible turbulence. For instance, a similar equation to equation \ref{eq:GRA_fluctallg} was derived by \cite{Rubesin1990}. He assumed that thermodynamic fluctuations behave in a polytropic manner:
\begin{equation}
	\frac{p'}{\ol{p}} = n \frac{\rho'}{\ol{\rho}} = \frac{n}{n-1} \; \frac{\rho T''}{\ol{\rho} \wt{T}} \qquad .
\end{equation}

\begin{figure}
\captionsetup[subfigure]{labelformat=empty}
\centering
\subfloat[]{\includegraphics[width=0.245\textwidth]{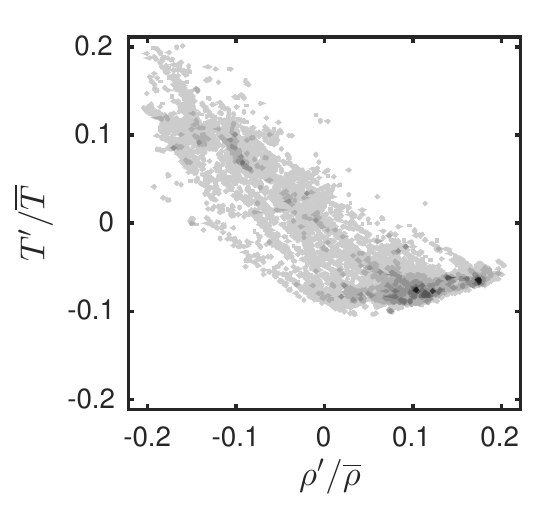}}
\subfloat[]{\includegraphics[width=0.245\textwidth]{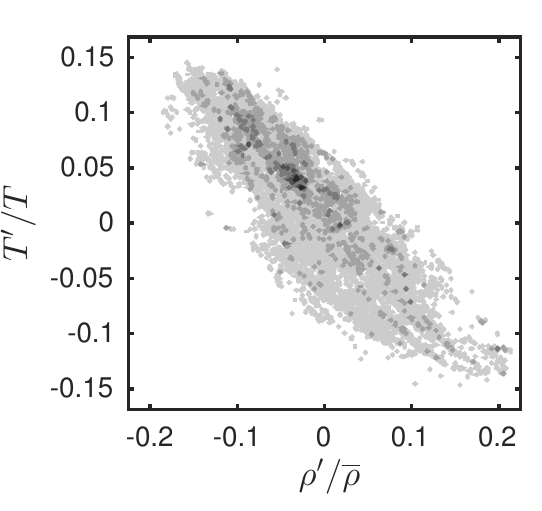}}
\subfloat[]{\includegraphics[width=0.245\textwidth]{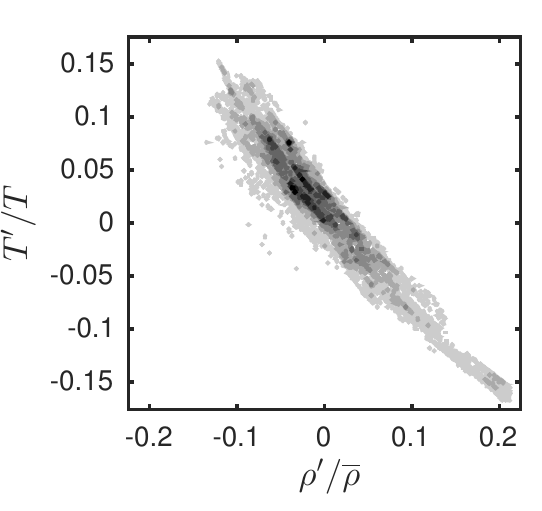}}
\subfloat[]{\includegraphics[width=0.245\textwidth]{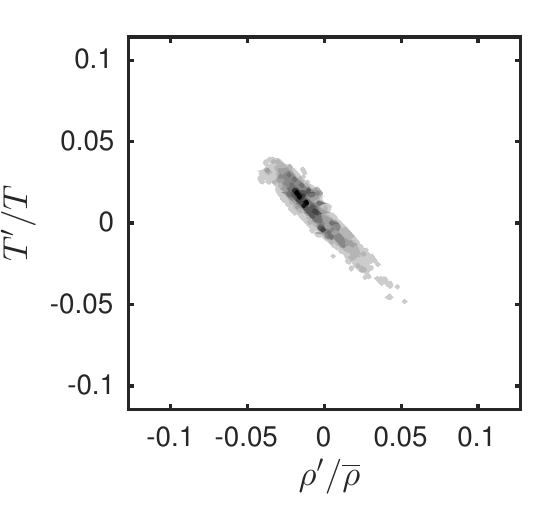}}\\[-8mm]

\subfloat[]{\includegraphics[width=0.245\textwidth]{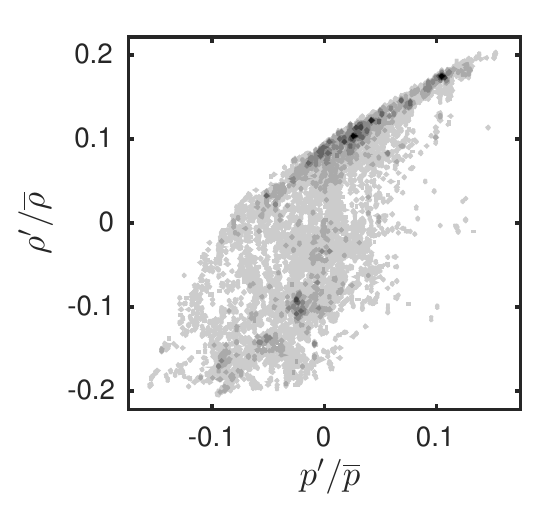}}
\subfloat[]{\includegraphics[width=0.245\textwidth]{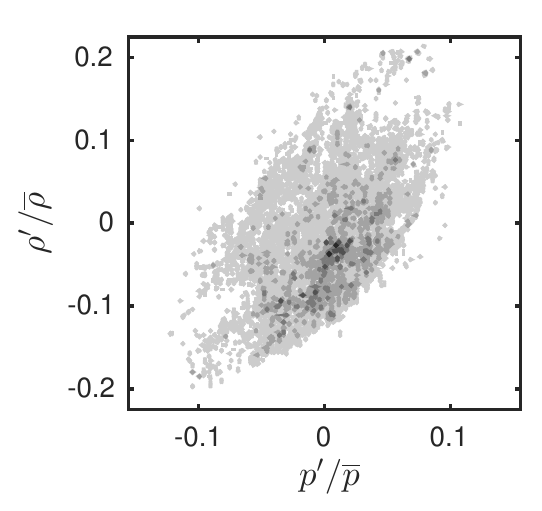}}
\subfloat[]{\includegraphics[width=0.245\textwidth]{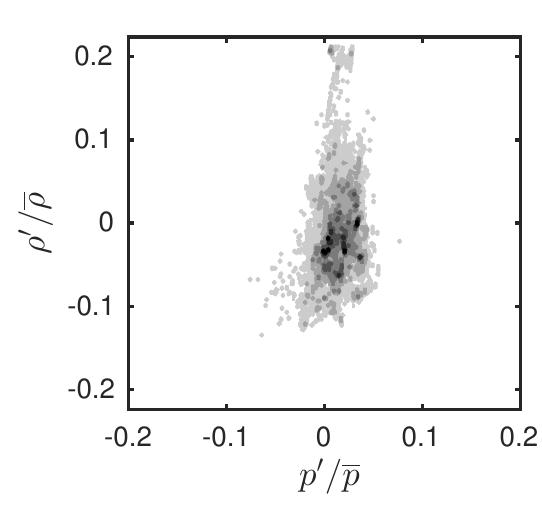}}
\subfloat[]{\includegraphics[width=0.245\textwidth]{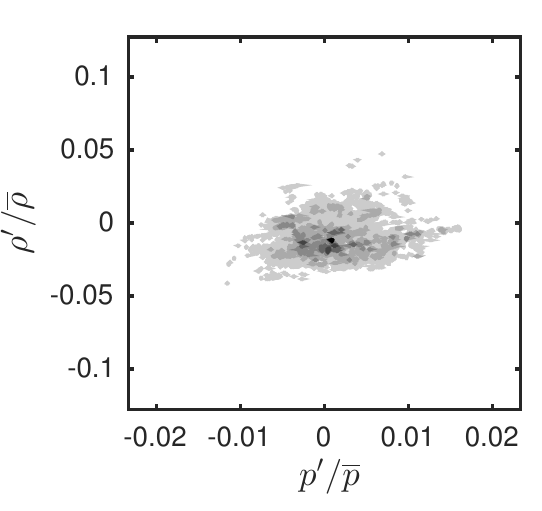}}\\[-8mm]

\subfloat[]{\includegraphics[width=0.245\textwidth]{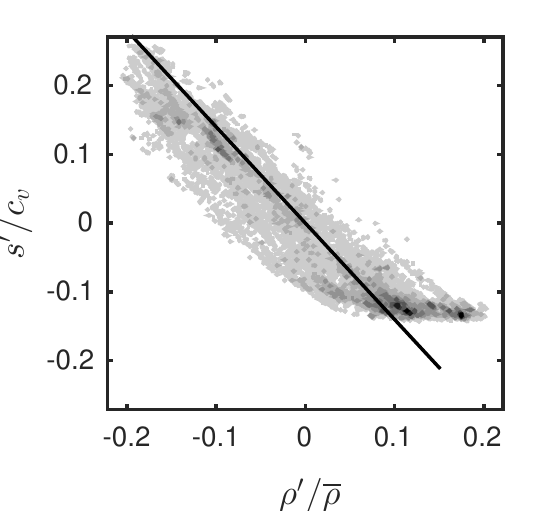}}
\subfloat[]{\includegraphics[width=0.245\textwidth]{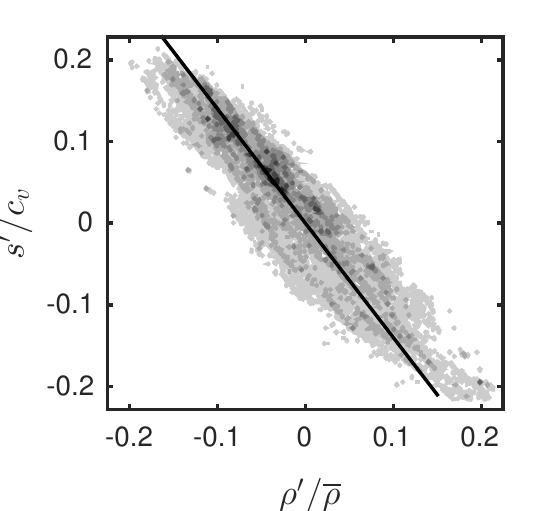}}
\subfloat[]{\includegraphics[width=0.245\textwidth]{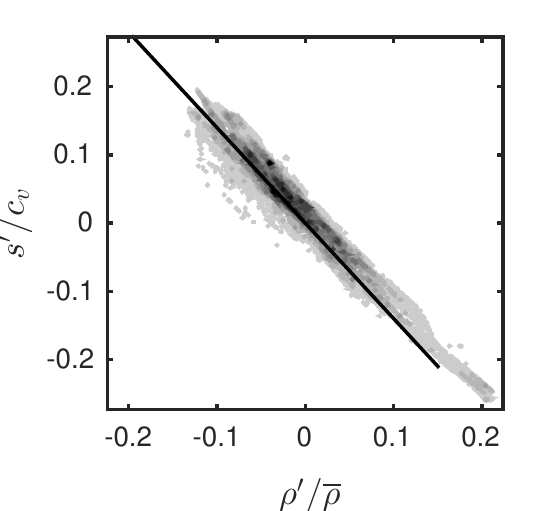}}
\subfloat[]{\includegraphics[width=0.245\textwidth]{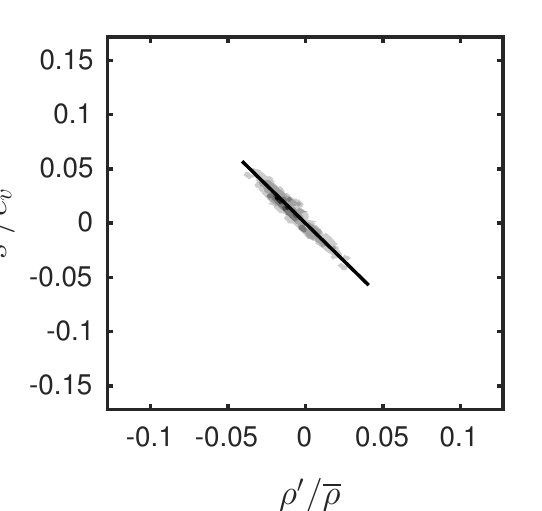}}\\[-8mm]

\subfloat[]{\includegraphics[width=0.245\textwidth]{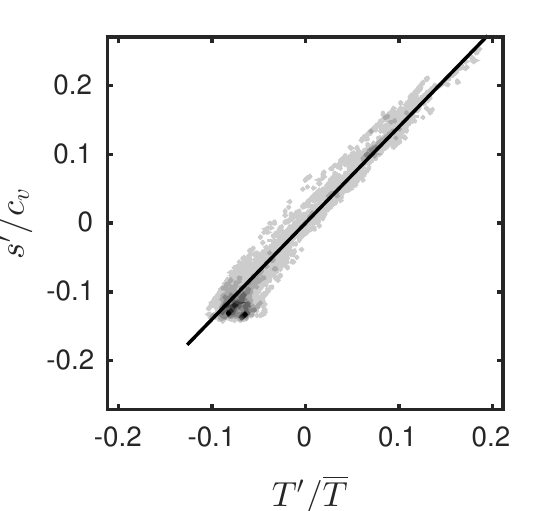}}
\subfloat[]{\includegraphics[width=0.245\textwidth]{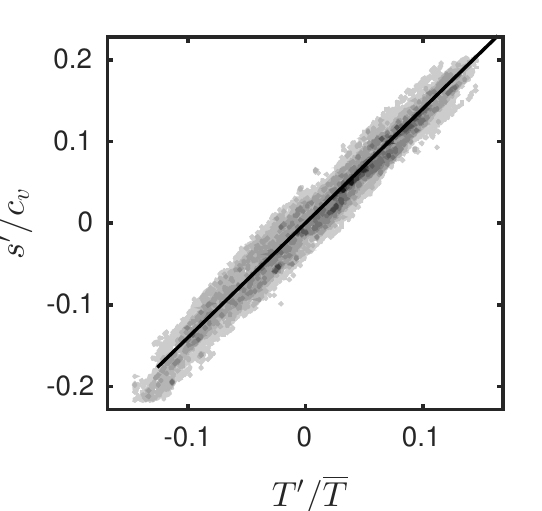}}
\subfloat[]{\includegraphics[width=0.245\textwidth]{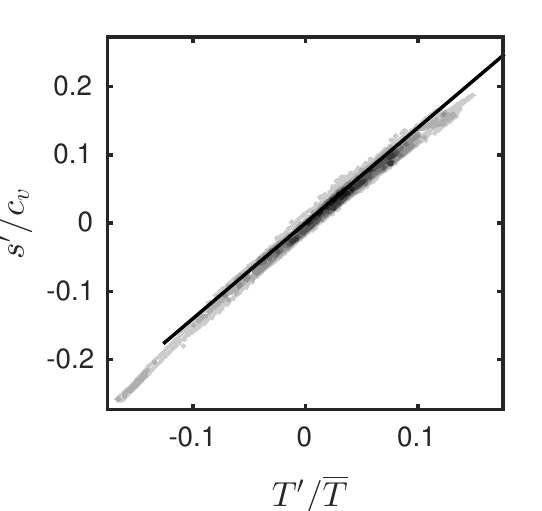}}
\subfloat[]{\includegraphics[width=0.245\textwidth]{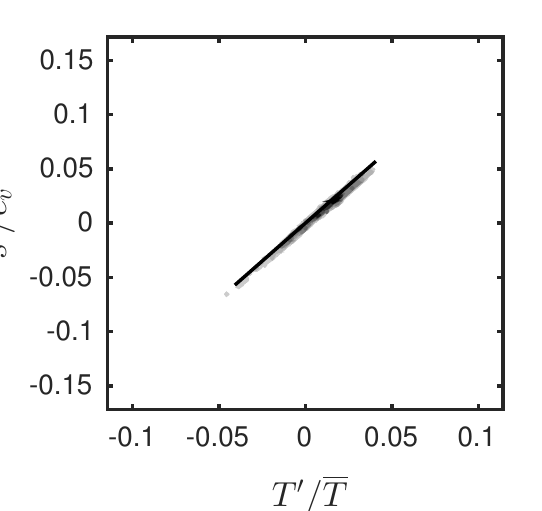}}\\[-8mm]

\subfloat[]{\includegraphics[width=0.245\textwidth]{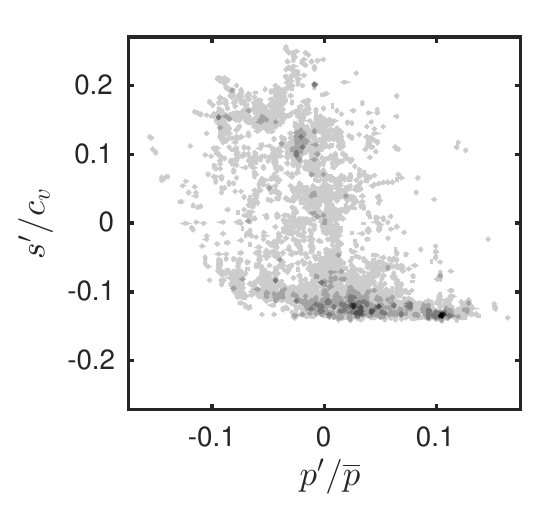}}
\subfloat[]{\includegraphics[width=0.245\textwidth]{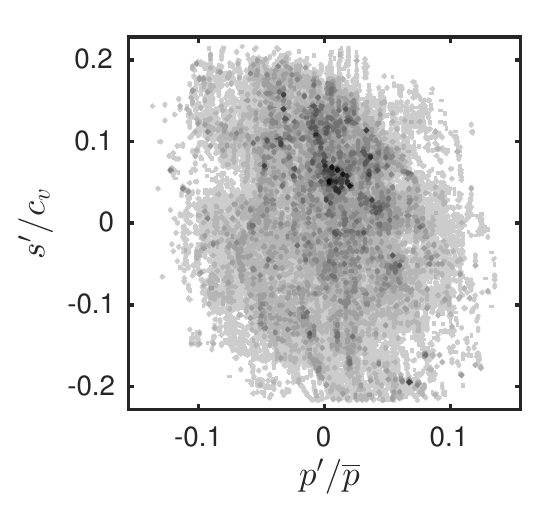}}
\subfloat[]{\includegraphics[width=0.245\textwidth]{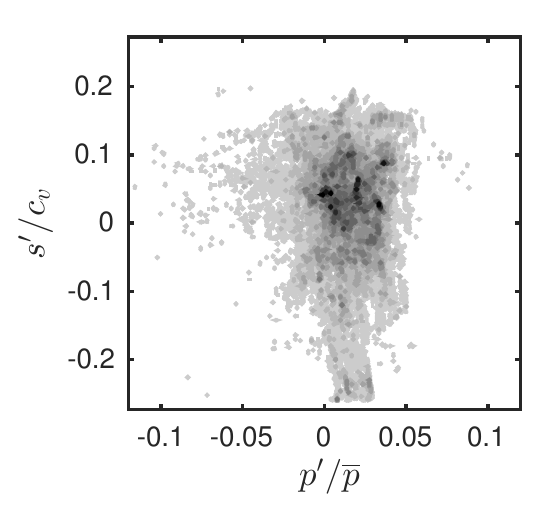}}
\subfloat[]{\includegraphics[width=0.245\textwidth]{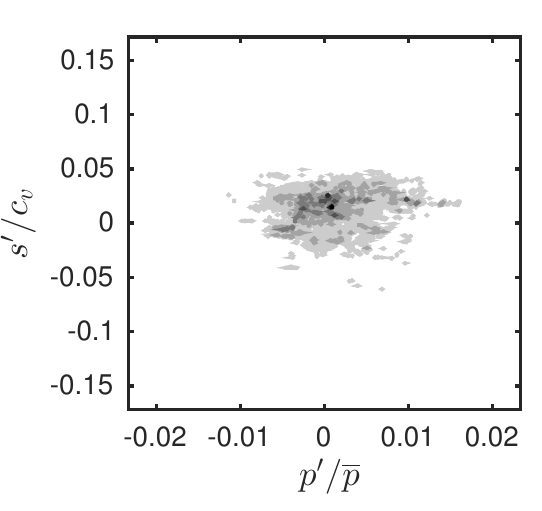}}\\[-8mm]

\caption{Scatter plots at $13 \leq y^+ \leq 17$ for different radial positions (left to right: $r/D=0.3$, $0.8$, $1.4$ and $3.5$, simulation \#6. In row 3 and 4 the approximation \ref{eq:corr_s} is included (black solid line).}
\label{fig:scat_6_yp15}
\end{figure}

\noindent After assuming that $T'/\ol{T} \approx T''/\wt{T}$ and linearisation according to \cite{LechnerSesterhenn2001} the relation
\begin{equation}
	\rk{n-1 }\frac{\rho'}{\ol{\rho}} \approx \frac{\rho T'}{\ol{\rho} \; \ol{T}}
\end{equation}

\noindent is derived. As suggested, with $n=0$ it follows that the correlation coefficient $R_{\rho,T}$ is minus one, $R_{\rho,p} \approx 0$ and that pressure fluctuations are unimportant compared to density fluctuations. The correlation coefficients are given in table \ref{tab:corrkoeff}, the scatterplots are shown in figure \ref{fig:scat_6_yp15}. $R_{\rho,T}$ is strongly negative for all observation points. Far away from the stagnation point ($r/D=3.5$), the coefficient reaches a value of $-0.94$ and justifies the approximations. The correlation between density and pressure is not zero, as proposed. On the contrary, the coefficient is strongly positive ($\approx 0.7$) close to the axis, but decreases with increasing $r/D$. Far away from the stagnation point, the approximation $R_{\rho,p} \approx 0$ is valid. Similarly, the pressure fluctuations are not unimportant compared to the density fluctuation in the entire region where the flow is influenced by the impingement. Farther downstream $\text{RMS}_{\rho}/\ol{\rho}$ is around three times as large as $\text{RMS}_p/\ol{p}$ for $y^+ \gtrsim 5$.

\begin{figure}
\captionsetup[subfigure]{labelformat=empty}
\centering
\subfloat[]{\includegraphics[width=0.245\textwidth]{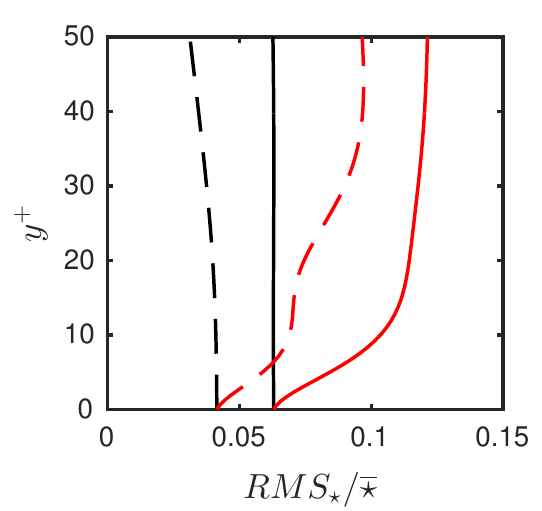}}
\subfloat[]{\includegraphics[width=0.245\textwidth]{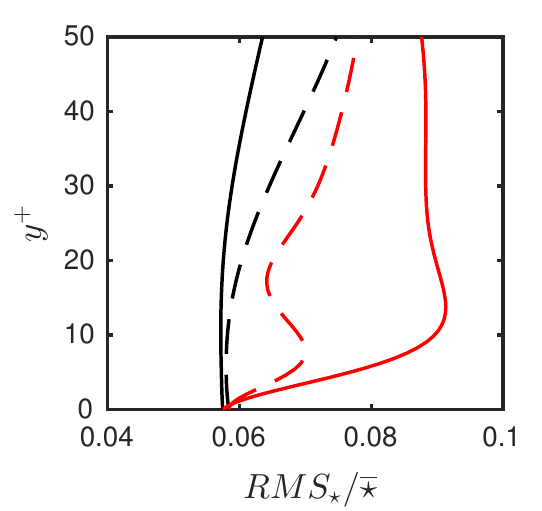}}
\subfloat[]{\includegraphics[width=0.245\textwidth]{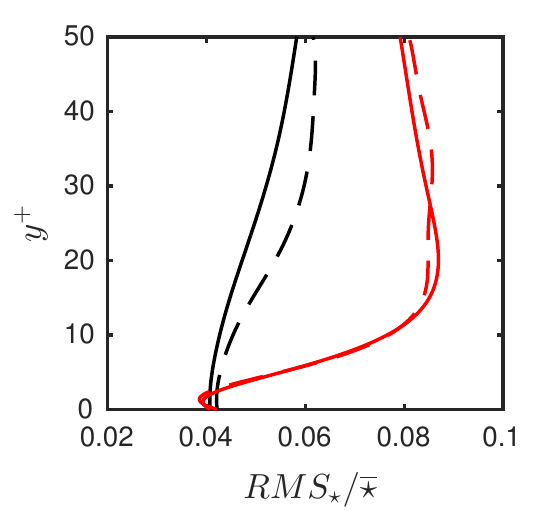}}
\subfloat[]{\includegraphics[width=0.245\textwidth]{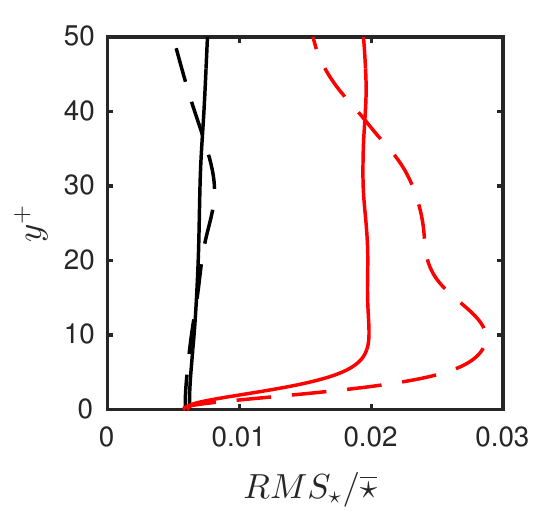}}\\[-8mm]
\caption{RMS values of pressure and density for sub-sonic impinging jets. \lindk: $\text{RMS}_p/\ol{p}$ (\#5), \link: $\text{RMS}_{p}/\ol{p}$ (\#6), \lindr: $\text{RMS}_{\rho}/\ol{\rho}$ (\#5), \linr: $\text{RMS}_{\rho}/\ol{\rho}$ (\#6)  }
\label{fig:RMS}
\end{figure}

\begin{table}
\centering
	\caption{Correlation coefficients for simulation \#6.}
	\begin{tabularx}{\columnwidth}{X r r r r r}
	\toprule
	$R$ & $y^+$ & $r/D=0.3 \pm 0.05$ & $r/D=0.8 \pm 0.05$ & $r/D=1.4 \pm 0.05$ & $r/D=3.5 \pm 0.05$\\
	\midrule
	$R_{v' u_r'}$	& $15 \pm 2$ & -0.04 &  0.01 &  0.04 &  0.08\\
	$R_{v' T'}$		& $15 \pm 2$ & -0.08 &  0.05 &  0.18 &  0.24\\
	$R_{T' \rho'}$	& $15 \pm 2$ & -0.75 & -0.80 & -0.86 & -0.94\\
	$R_{\rho' p'}$	& $15 \pm 2$ &  0.70 &  0.58 &  0.41 &  0.14\\

	\midrule
	$R_{v' u_r'}$	& $40 \pm 2$ & -0.11 & -0.11 &  0.19 &  0.18\\
	$R_{v' T'}$		& $40 \pm 2$ & -0.02 &  0.11 & -0.11 & -0.24\\
	$R_{T' \rho'}$	& $40 \pm 2$ & -0.79 & -0.76 & -0.75 & -0.94\\
	$R_{\rho' p'}$	& $40 \pm 2$ &  0.69 &  0.40 &  0.48 &  0.04\\
	\end{tabularx}
	\label{tab:corrkoeff}
\end{table}

\cite{LechnerSesterhenn2001} assumed a linear relation between thermodynamic fluctuations. Following the entropy definition $s=c_v \text{ln}\rk{p/\rho^{\kappa}}$, the linearised gas law and the neglect of pressure fluctuations with respect to density fluctuations, the approximation reads:
\begin{equation}
	\frac{s'}{c_v} \approx - \kappa \frac{\rho'}{\ol{\rho}} \approx \kappa \frac{T'}{\ol{T}} \qquad .
	\label{eq:corr_s}
\end{equation}

\noindent Figure \ref{fig:scat_6_yp15} shows the scatter plots of ($s',\rho'$), ($s',T'$) and ($s',p'$). Approximation \ref{eq:corr_s} is included in the plots (black solid line) and can be confirmed for all radial positions.

\section{Turbulent heat flux}

\begin{figure}
\captionsetup[subfigure]{labelformat=empty}
\centering
\subfloat[]{\includegraphics[width=0.32\textwidth]{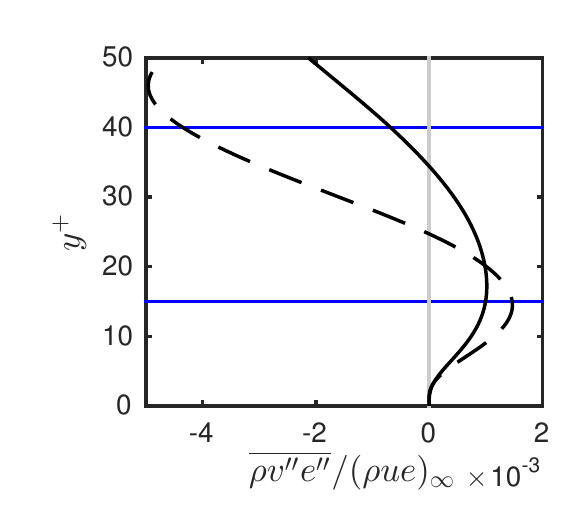}}
\subfloat[]{\includegraphics[width=0.32\textwidth]{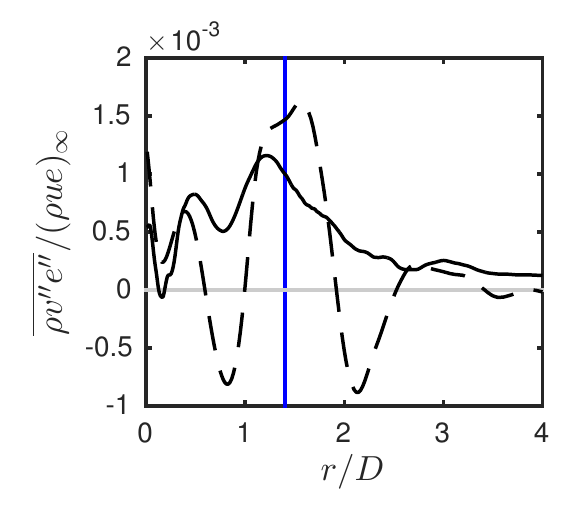}}
\subfloat[]{\includegraphics[width=0.32\textwidth]{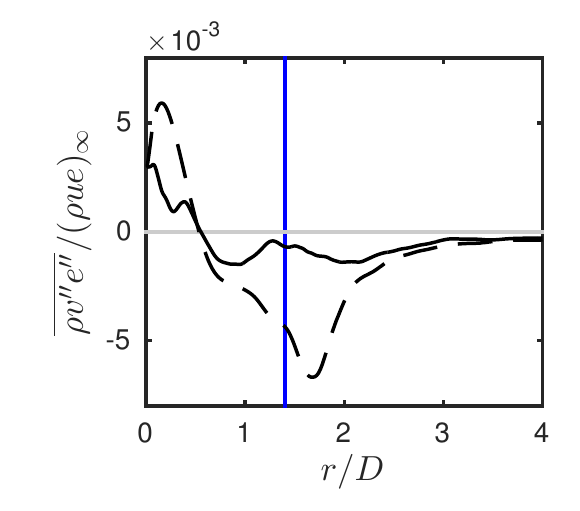}}\\[-8mm]

\subfloat[]{\includegraphics[width=0.32\textwidth]{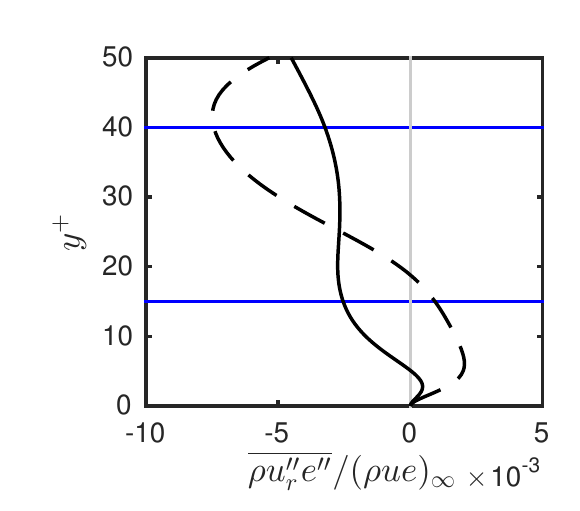}}
\subfloat[]{\includegraphics[width=0.32\textwidth]{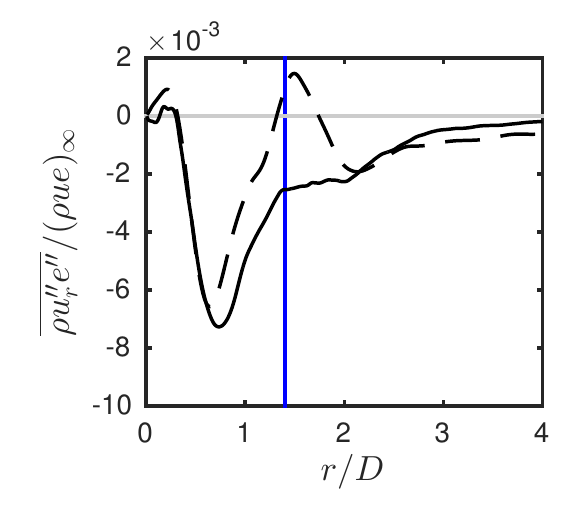}}
\subfloat[]{\includegraphics[width=0.32\textwidth]{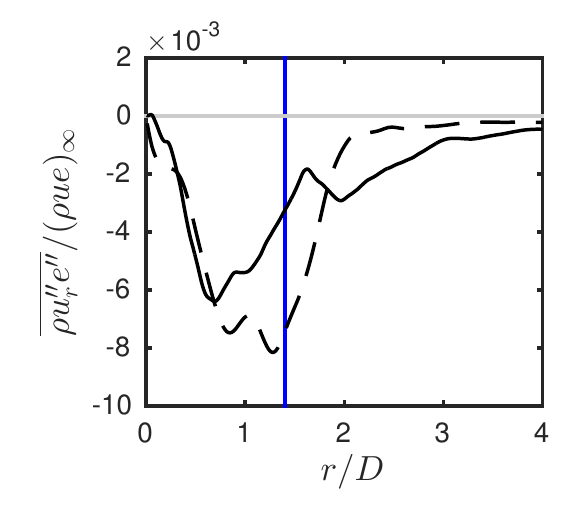}}\\[-6mm]

\caption{Influence of the Reynolds number (\#5: 3300, \#6: 8000) on the turbulent heat flux in wall normal direction $\ol{\rho v'' e''}$  (first row) and radial direction $\ol{\rho u_r'' e''}$ (second row). Left: $r/D=1.4$, middle: $y^+=15$, right: $y^+=40$. \linfuenf: \#5, \linsechs: \#6. For plots at fixed $r/D$, the horizontal locations $y^+=15$ and $y^+=40$ are shown (\linb). For plots at fixed $y^+$, the vertical position $r/D=1.4$ is shown (\linb).}
\label{fig:THF}
\end{figure}

In section \ref{sec:RA_mean}, figure \ref{fig:Cf_Nu} (e,f) the Nusselt number profiles of the simulations \#5 ($\Rey=3300$) and \#6 ($\Rey=8000$) are shown. In both cases, $\Nus$ decreases with increasing radial distance. Superimposed to this main trend, a shoulder exists at $1 \lesssim r/D \lesssim 2$, where the slope is reduced. Depending on the parameters of the impinging jet, the slope can also be positive in this area and form a secondary maximum. In \cite{WilkeSesterhenn2014} this shoulder or secondary maximum was ascribed to the occurrence of secondary vortex rings that locally increase the heat transfer. Using the turbulent heat flux, we can quantify this effect. The first row of figure \ref{fig:THF} shows the turbulent heat flux in wall normal direction $\ol{\rho v'' e''}$. Close to the wall, $\ol{\rho v'' e''}$ is positive in the area of the $\Nus$-shoulder (left plot, $r/D=1.4$). This means that the heat is transported in positive $y$-direction (away from the impinging plate). After reaching a maximum at $y^+ \approx$ 15 to 20, the heat flux decreases and turns negative. This is due to the fact, that vortices present at the upper border of the wall jet entrain hot fluid. The radial distribution at $y^+=15$ (middle plot) proves that the zone, where the turbulent heat flux is strongly positive coincides with the $\Nus$-shoulder and that $\ol{\rho v'' e''}$ is negative in the vicinity of this zone for the lower Reynolds number. Further can be determined that, in this zone, the influence of $\ol{\rho v'' e''}$ is weaker at the higher Reynolds number. This agrees with the fact that the $\Nus$-profile is smoother and that the vortex rings are overlaid with small scale turbulence, leading to less strong structures.

The second row of figure \ref{fig:THF} shows the turbulent heat transfer in radial direction $\ol{\rho u_r'' e''}$. It is of the same order of magnitude like the one in wall normal direction. Close to the wall, $\ol{\rho v'' e''}$ is positive at $r/D=1.4$ for the case of $\Rey=3300$. For the higher Reynolds number, this area of downstream turbulent heat flux is almost not present. At larger wall distances, $\ol{\rho v'' e''}$ is negative for both cases due to the direction of rotation of the vortices in the shear layers.

\section{Reynolds stresses}

\begin{figure}
\captionsetup[subfigure]{labelformat=empty}
\centering
\subfloat[]{\includegraphics[width=0.32\textwidth]{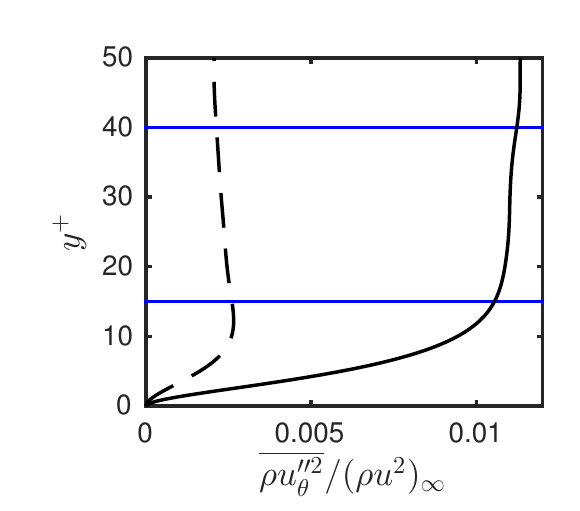}}
\subfloat[]{\includegraphics[width=0.32\textwidth]{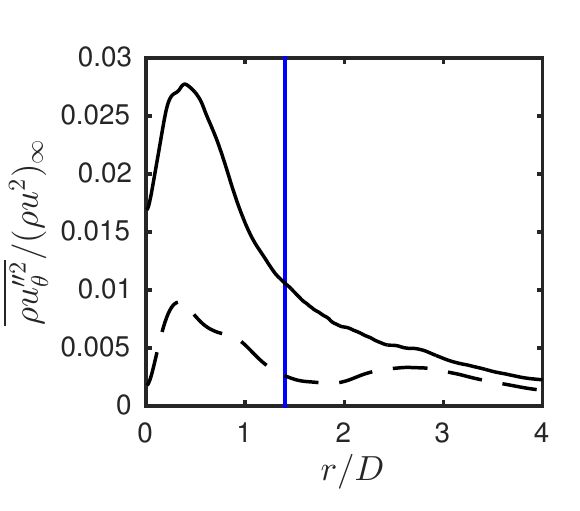}}
\subfloat[]{\includegraphics[width=0.32\textwidth]{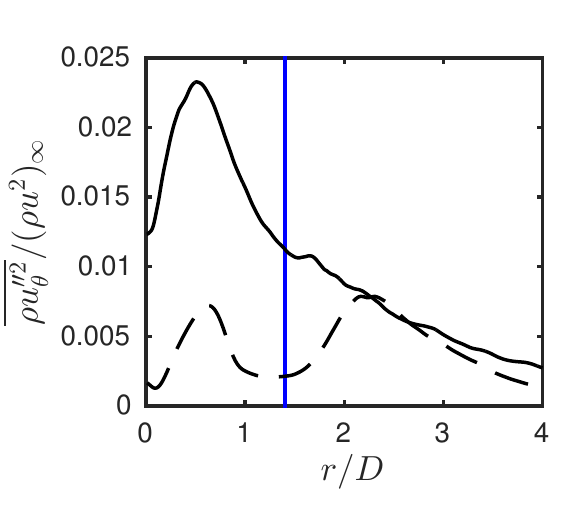}}\\[-8mm]

\subfloat[]{\includegraphics[width=0.32\textwidth]{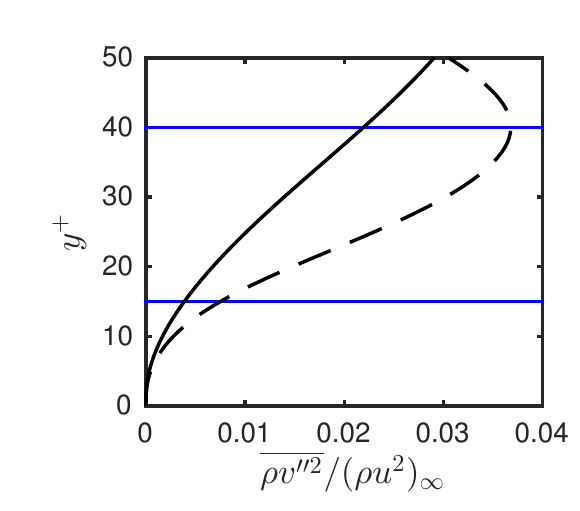}}
\subfloat[]{\includegraphics[width=0.32\textwidth]{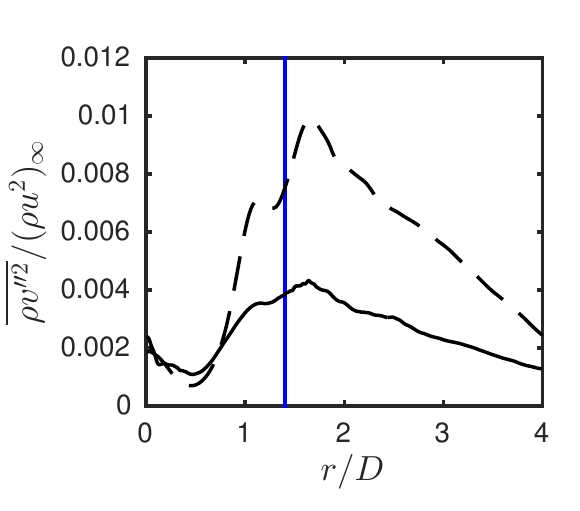}}
\subfloat[]{\includegraphics[width=0.32\textwidth]{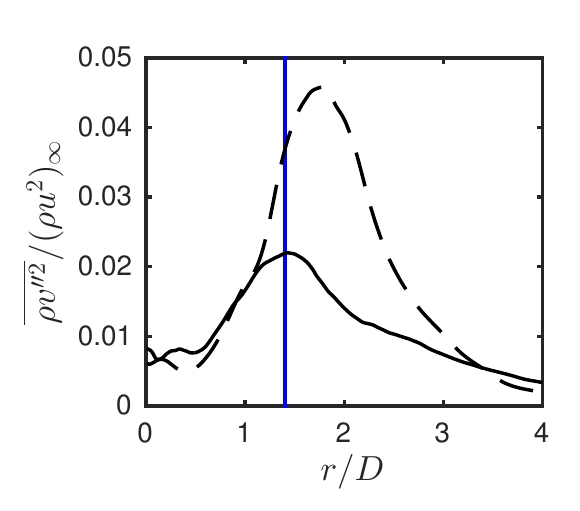}}\\[-8mm]

\subfloat[]{\includegraphics[width=0.32\textwidth]{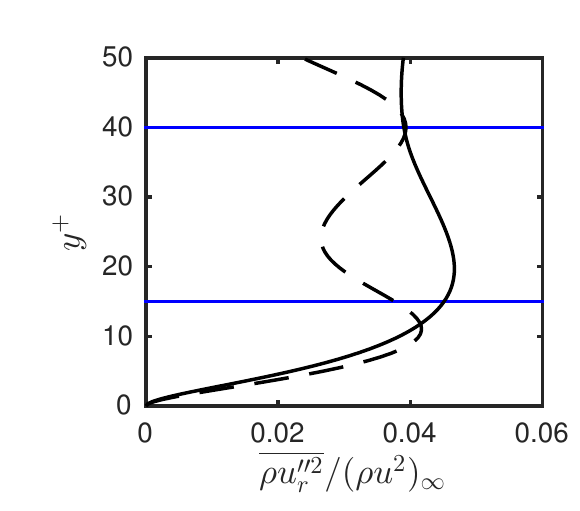}}
\subfloat[]{\includegraphics[width=0.32\textwidth]{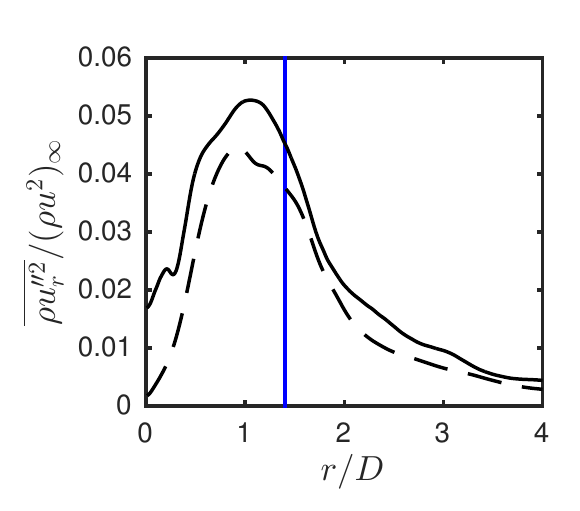}}
\subfloat[]{\includegraphics[width=0.32\textwidth]{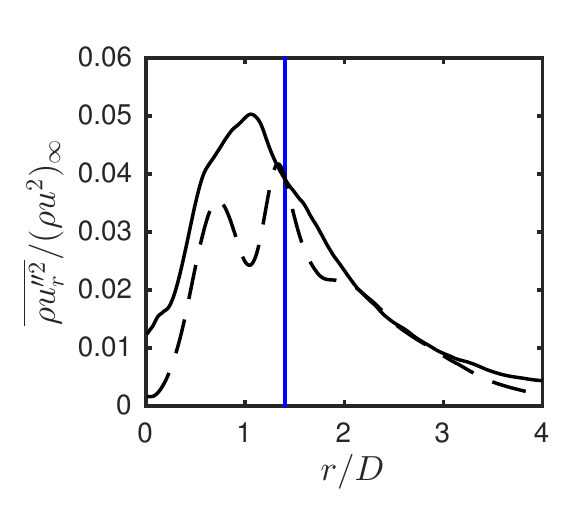}}\\[-8mm]

\subfloat[]{\includegraphics[width=0.32\textwidth]{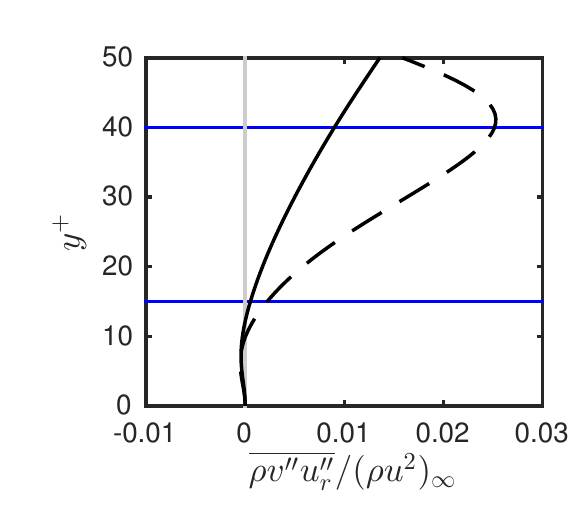}}
\subfloat[]{\includegraphics[width=0.32\textwidth]{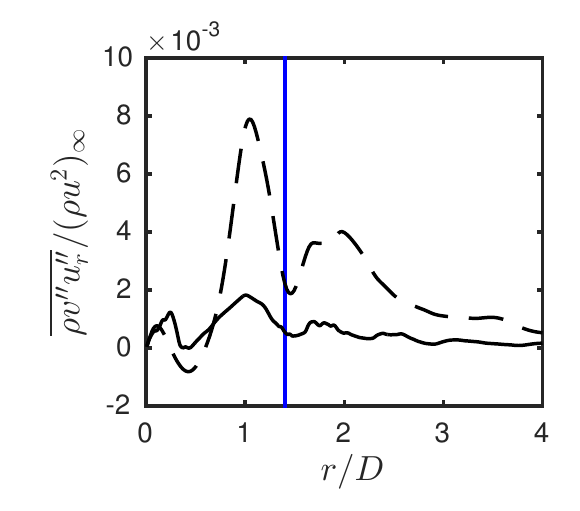}}
\subfloat[]{\includegraphics[width=0.32\textwidth]{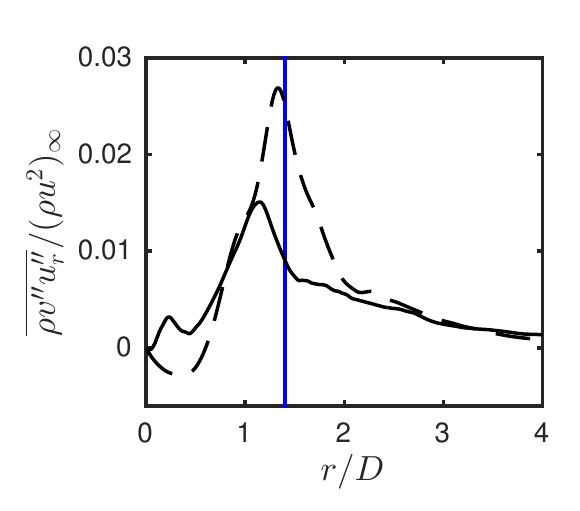}}\\[-6mm]

\caption{Influence of the Reynolds number (\#5: 3300, \#6: 8000) on the main Reynolds stress tensor components, top to down: $\ol{\rho u_{\theta}''^2}$, $\ol{\rho v''^2}$, $\ol{\rho u_r''^2}$ and $\ol{\rho v'' u_r''}$. Left: $r/D=1.4$, middle: $y^+=15$, right: $y^+=40$. \linfuenf: \#5, \linsechs: \#6. For plots at fixed $r/D$, the horizontal locations $y^+=15$ and $y^+=40$ are shown (\linb). For plots at fixed $y^+$, the vertical position $r/D=1.4$ is shown (\linb).}
\label{fig:RST}
\end{figure}

Modern high performance computers allow three-dimensional direct numerical simulations of impinging jets with relevant Reynolds numbers since recently. However, the computations are limited to academic cases for the foreseeable future. For the improvement of turbulence models, that are widely used for industrial applications, our DNS provide a database to compare with. An important term that rises among others in the Reynolds-averaged Navier-Stokes equations (RANS) is the Reynolds stress tensor $\ol{\rho u_i'' u_j''}$, that we describe in this section. Since the mean circumferential velocity component is zero, the terms $\ol{\rho u_{\theta}'' v''}$ and $\ol{\rho u_{\theta}'' u_r''}$ are not of relevance and therefore not shown. Figure \ref{fig:RST} shows the four main entries of the tensor. We analyse the flow close to the wall and at the the radial distance, where the shoulder of the Nusselt number is present. The profiles are taken normal to the wall at $r/D=1.4$ (left column) and parallel to the wall at $y^+=15$ (middle column) and $y^+=40$ (right column). 

At $r/D=1.4$, the $\theta\theta$-component (first row, left) increases until $y^+ \approx 10$ and then stays almost constant within the boundary layer. The maximum stress occurs around a half diameter from the jet axis. At low Reynolds number (3300) a second maximum is present around $r/D = 2 - 2.5$. Another difference between the two simulations is that the stress component is higher in the case of the higher Reynolds number (8000). Additional to this area another area of high stress is present farther away from the wall and the jet axis and will be discussed later.

The $yy$-component is stronger in the case of the low Reynolds number. The strength of this entry of the tensor increases with the distance to the wall, until it decreases again when approaching the upper end of the wall jet. Since the wall jet's center and upper end is at higher $y^+$-values in the case of the higher Reynolds number, the maximum of $\ol{\rho v''^2}$ occurs also at a higher dimensionless wall distance. The highest stress occurs at a radial distance of around $1.5 - 1.8$ diameters.

\begin{figure}
\captionsetup[subfigure]{labelformat=empty}
\centering
\subfloat[]{\includegraphics[width=0.495\textwidth]{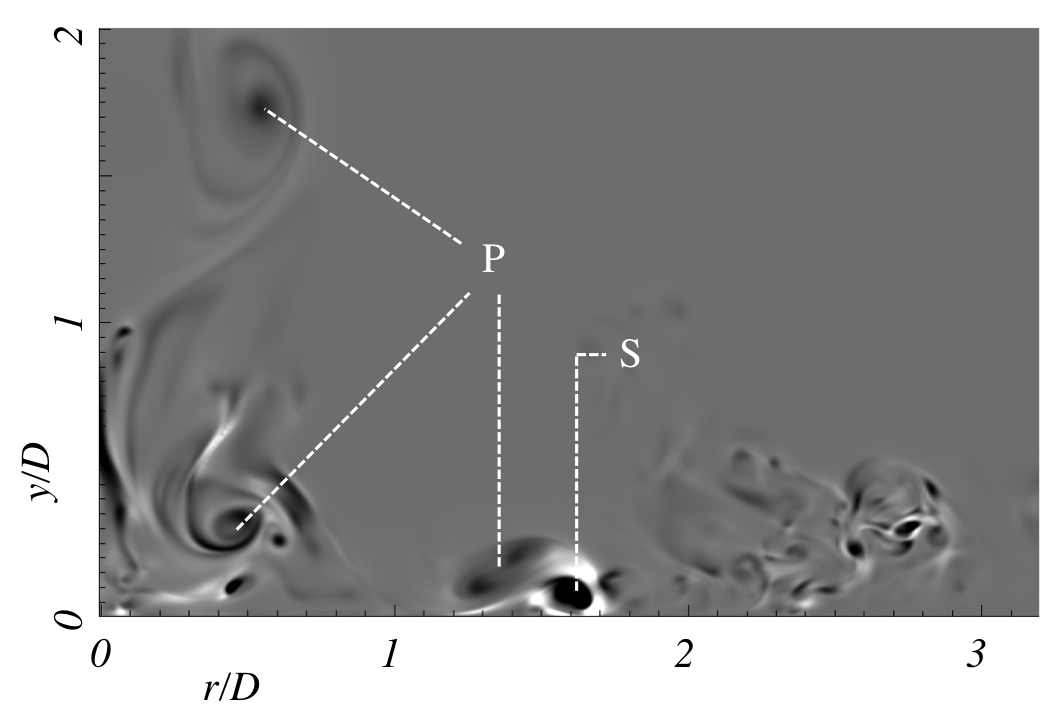}}
\subfloat[]{\includegraphics[width=0.495\textwidth]{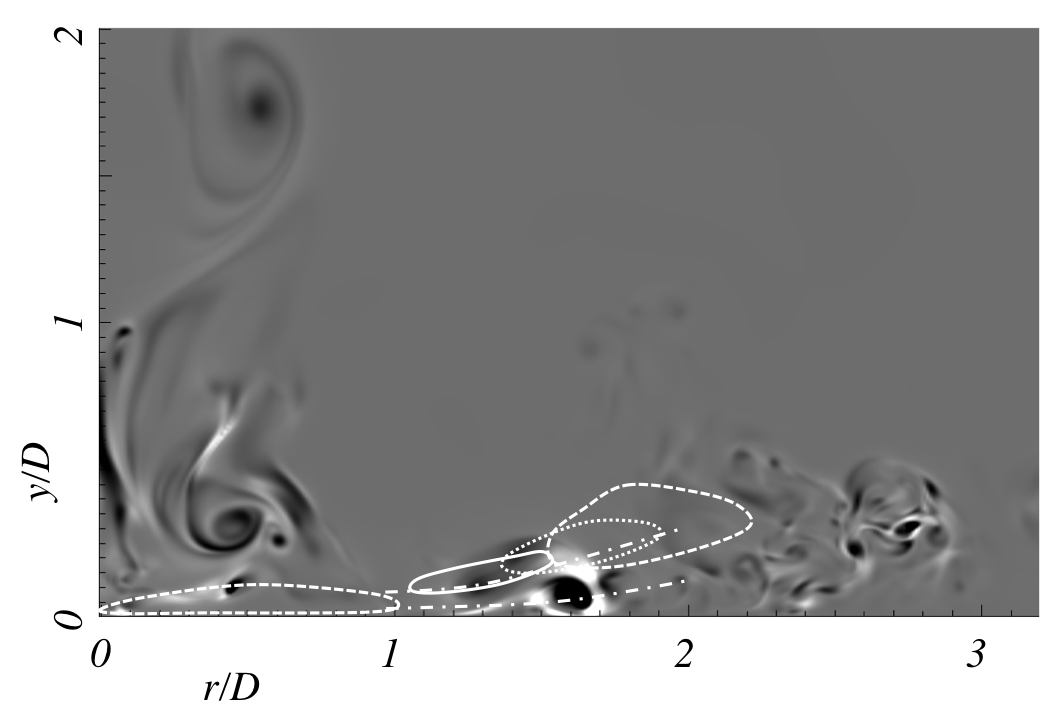}}\\[-8mm]

\caption{Instantaneous flow field of simulation \#6: $Q D^2 / u_{\infty}^2$ in the range $-85$ (white) to $85$ (black). Left: location of primary (P) and secondary (S) vortices. Right: locations of high Reynolds stress tensor components \linOdk: $\theta\theta$, \linOpk: $yy$, \linOddk: $rr$, \link: $yr$}
\label{fig:RST_skizze}
\end{figure}

The distribution of the $rr$-component features two maxima in wall normal direction, whereby this characteristic is distinct stronger in the case of the lower Reynolds number. The gap between the two maxima increases with the distance from the jet axis. This shape is directly caused by the primary and secondary vortices that move parallel to the wall and increase their diameter. In figure \ref{fig:RST_skizze}, the area of the highest $\ol{\rho u_r''^2}$-value is indicated by two dash-dotted lines. Those lines coincide also with the path of the primary vortex rings that have their origin in the shear layer of the free jet and the counter-rotating secondary vortices that emerge due to the wall-friction. The maximal fluctuations occur at $r/D \approx 1$, where the two vortices have the highest radial velocity.

The $yr$-component has a similar wall-normal distribution like the $yy$-component. However, the radial location of the highest value is closer to the jet axis at $r/D \approx 1 - 1.3$. The simulation with the lower Reynolds number features higher values of $\ol{\rho v'' u_r''}$ and a stronger distinct maximum.

Figure \ref{fig:RST_skizze} summarises the locations of high stresses. The impingement of the primary ring vortices cause strong stress in $\theta\theta$-direction. The movement of the pair consisting of a primary and a secondary vortex causes strong stress in $rr$-direction. The components $yr$, $yy$ and again $\theta\theta$ are strongly influenced by the movement of the primary vortex and become important in this order with increasing radial distance.

\section{Reynolds stress budgets}

Performing DNS, we are able to compute the terms in the balance equations for the Reynolds stress tensor components. Since the mean flow of the impinging jet is axisymmetrical, we use a cylindrical coordinate system for the analysis. In \cite{MoserMoin1984} the balance equations are derived for cylindrical coordinates using only Reynolds-averages. For compressible flows it is common practice to use the Reynolds- (mean: $\ol{\star}$, fluctuation: $\star'$) and Favre-average (mean: $\wt{\star}$, fluctuation: $\star''$) simultaneously. After statistical averaging of the Navier-Stokes equations and some transformations, we derived the Reynolds stress transport equation in the following form:

\begin{align}
	\dt{\ol{\rho u_i'' u_j''}} &+ \ub{\pd{}{x_k}\rk{\wt{u_k} \ol{\rho u_i'' u_j''}}}_{C_{ij}} = \ub{- \ol{\rho u_{i}'' u_{k}''}\pd{\wt{u_{j}}}{x_k} - \ol{\rho u_{j}'' u_{k}''}\pd{\wt{u_{i}}}{x_k}}_{PR_{ij}}\nn\\
	& \ub{+ \pd{}{x_k}\left[\ol{\rho u_i'' u_j'' u_k''} + \ol{p'\left(u_i' \delta_{jk} + u_j' \delta_{ik}\right)}\right]}_{TD_{ij}}  \ub{- \ol{\tau_{ik}' \pd{u_{j}'}{x_k}} - \ol{\tau_{jk}' \pd{u_{i}'}{x_k}}}_{DS_{ij}} \nn\\
	& \ub{+ \pd{}{x_k}\left(\ol{u_i' \tau_{jk}'} + \ol{u_j' \tau_{ik}'}\right)}_{VD_{ij}} \ub{+ \ol{p'\left(\pd{u_{i}'}{x_j}+\pd{u_j'}{x_i}\right)}}_{PS_{ij}} \nn\\
	& \ub{+ \ol{u_i''}\left(\pd{\ol{\tau_{jk}}}{x_k} -\pd{\ol{p}}{x_j}\right) + \ol{u_j''}\left(\pd{\ol{\tau_{ik}}}{x_k} -\pd{\ol{p}}{x_i}\right)}_{M_{ij}} \qquad .
\end{align}

\noindent The terms mean the following: $C$: convection, $PR$: production, $TD$: turbulent diffusion, $DS$: turbulent dissipation, $VD$: viscous diffusion, $M$: mass-flux variation and $PS$: pressure-strain. The imbalance of the budget is denoted $IB$. The expressions for the regarded components of cylindrical coordinates are given in appendix \ref{appA}. The averaging was performed in time as well as in circumferential direction. Therefore all derivatives with respect to $\theta$ are zero. Despite the circumferential velocity is small $u_{\theta}/u_{\infty} \lesssim 0.01$, it was not neglected. The budgets are presented for one radial position $r/D=1.4$ and at one constant dimensionless wall distance $y^+=15$.

\begin{figure}
\captionsetup[subfigure]{labelformat=empty}
\centering
\subfloat[]{\includegraphics[width=0.32\textwidth]{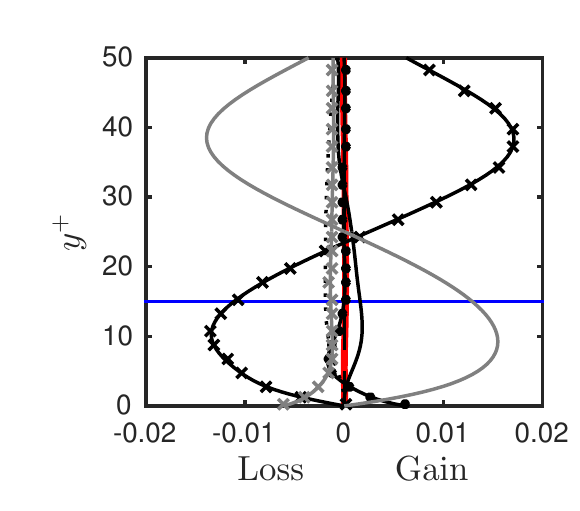}}
\subfloat[]{\includegraphics[width=0.32\textwidth]{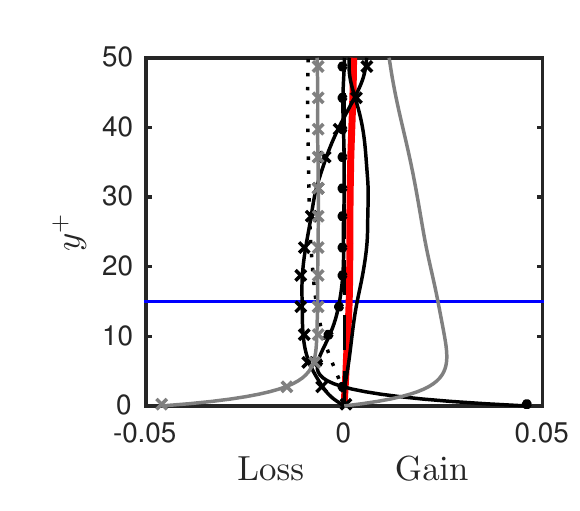}}\\[-10mm]

\subfloat[]{\includegraphics[width=0.32\textwidth]{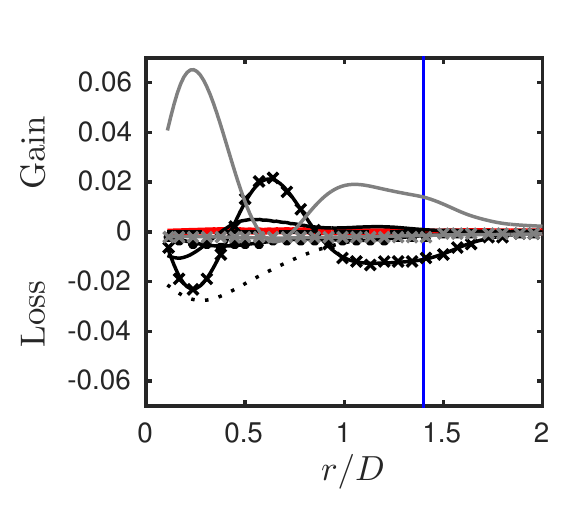}}
\subfloat[]{\includegraphics[width=0.32\textwidth]{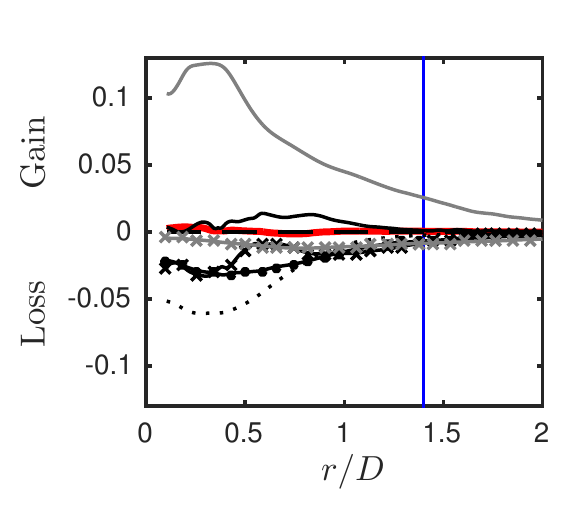}}\\[-8mm]
\caption{Budget of the $\theta\theta$-component $\ol{\rho u_{\theta}''^2}$. Left column: \#5 ($\Rey=3300$), right colomn: \#6 ($\Rey=8000$), first row: $r/D=1.4$, second row: $y^+=15$. \budGGW: $IB$, \budC: $C$, \budPR: $PR$, \budTD: $TD$, \budVD: $VD$, \budM: $M$, \budPS: $PS$, \budDS: $DS$. For plots at fixed $r/D$, the horizontal location $y^+=15$ is shown (\linb). For plots at fixed $y^+$, the vertical position $r/D=1.4$ is shown (\linb).}
\label{fig:RST_bud_th2}
\end{figure}

The Budget of $\ol{\rho u_{\theta}''^2}$ in figure \ref{fig:RST_bud_th2} shows significant differences between the two different Reynolds numbers. At $r/D=1.4$, the two dominant terms in the boundary layer, but not at the wall are turbulent diffusion and pressure strain in simulation \#5 ($\Rey=3300$). Both are of equal strength and opposite sign. At the higher Reynolds number, the turbulent diffusion is of less importance. As counterpart to the pressure strain, also production and turbulent dissipation contribute to the loss of stress. At the wall, stress is produced by viscous diffusion and lost by turbulent dissipation in both cases.
At $y^+=15$, stress is mainly produced by pressure strain and lost by convection (both cases). At $\Rey=3300$, turbulent diffusion contributes positively at $r/D \approx 0.6$ and negatively at other radial distances. In contrary, the turbulent diffusion is negative for $\Rey=8000$ for all radial distances.

\begin{figure}
\captionsetup[subfigure]{labelformat=empty}
\centering
\subfloat[]{\includegraphics[width=0.32\textwidth]{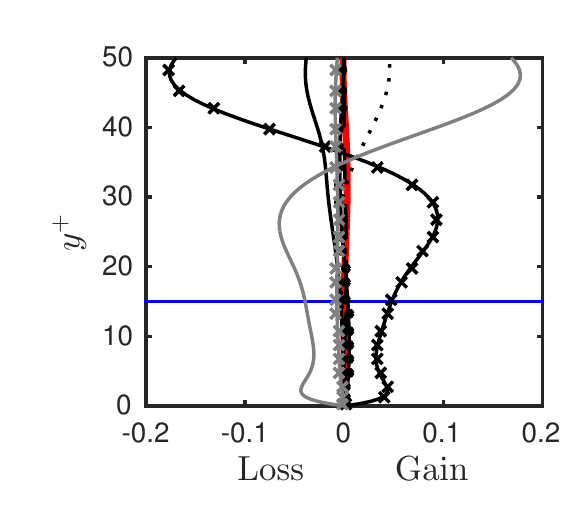}}
\subfloat[]{\includegraphics[width=0.32\textwidth]{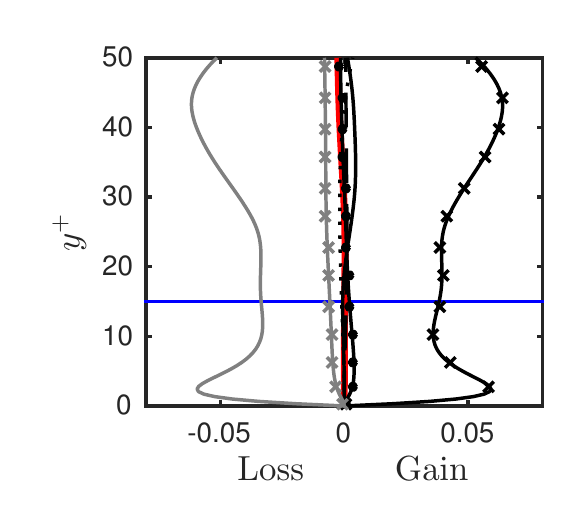}}\\[-10mm]

\subfloat[]{\includegraphics[width=0.32\textwidth]{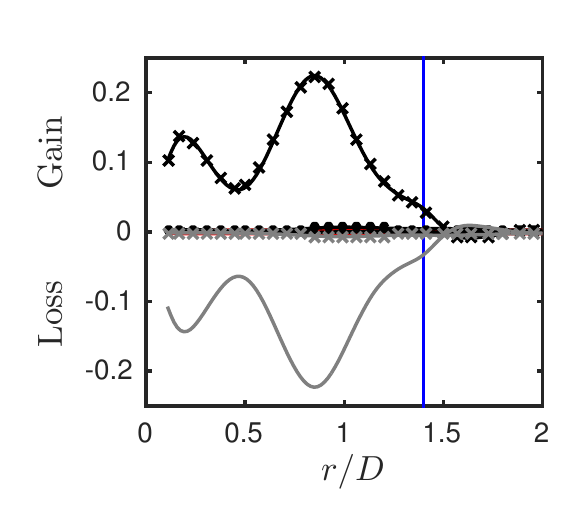}}
\subfloat[]{\includegraphics[width=0.32\textwidth]{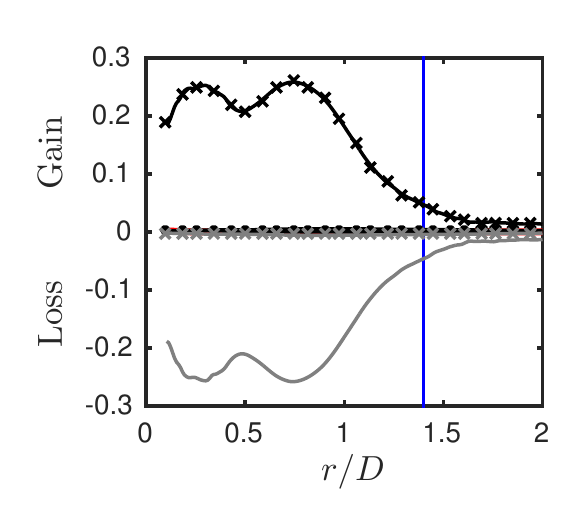}}\\[-8mm]
\caption{Budget of the $yy$-component $\ol{\rho v''^2}$. Left column: \#5 ($\Rey=3300$), right colomn: \#6 ($\Rey=8000$), first row: $r/D=1.4$, second row: $y^+=15$. \budGGW: $IB$, \budC: $C$, \budPR: $PR$, \budTD: $TD$, \budVD: $VD$, \budM: $M$, \budPS: $PS$, \budDS: $DS$, \linb: locations $r/D=1.4$, $y^+=15$.}
\label{fig:RST_bud_y2}
\end{figure}

The budget of $\ol{\rho v''^2}$ in figure \ref{fig:RST_bud_y2} is dominated by turbulent diffusion and pressure strain in both cases and all locations. Except for $y^ +\gtrsim 35$ at $\Rey=3300$, $TD$ contributes positively and $PS$ negatively. At this higher dimensionless wall distances, also convection and production become more important and contribute with loss respectively gain. In radial direction, the two opponents feature maxima at $r/D \approx 0.25$ and $r/D \approx 0.8$.

\begin{figure}
\captionsetup[subfigure]{labelformat=empty}
\centering
\subfloat[]{\includegraphics[width=0.32\textwidth]{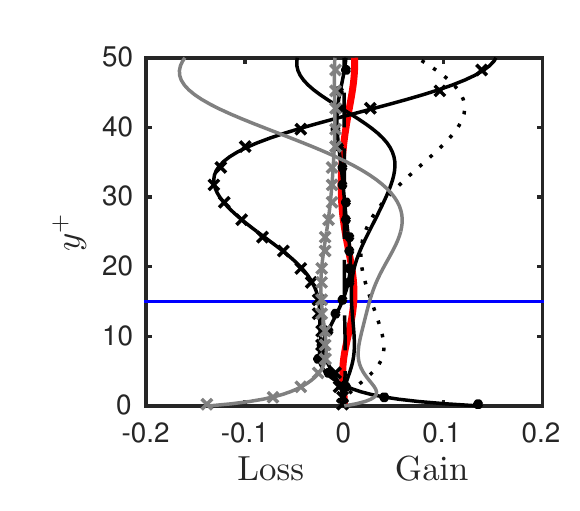}}
\subfloat[]{\includegraphics[width=0.32\textwidth]{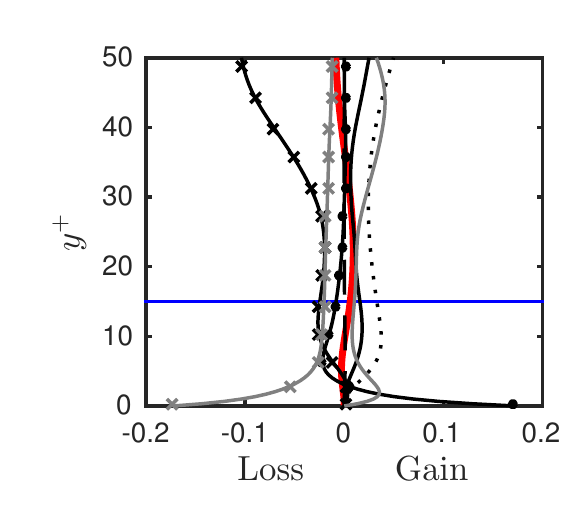}}\\[-10mm]

\subfloat[]{\includegraphics[width=0.32\textwidth]{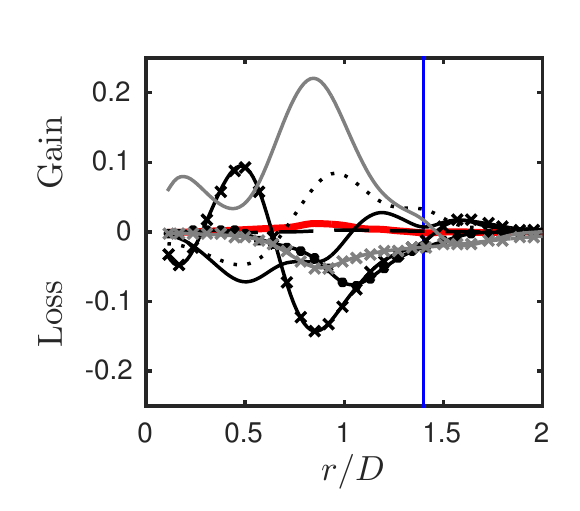}}
\subfloat[]{\includegraphics[width=0.32\textwidth]{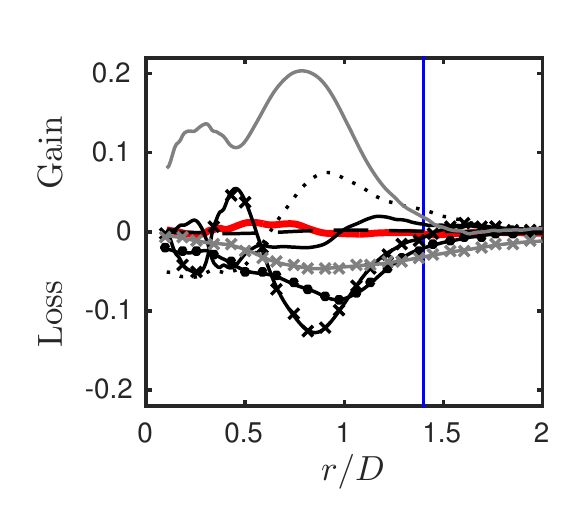}}\\[-8mm]
\caption{Budget of the $rr$-component $\ol{\rho u_r''^2}$. Left column: \#5 ($\Rey=3300$), right colomn: \#6 ($\Rey=8000$), first row: $r/D=1.4$, second row: $y^+=15$. \budGGW: $IB$, \budC: $C$, \budPR: $PR$, \budTD: $TD$, \budVD: $VD$, \budM: $M$, \budPS: $PS$, \budDS: $DS$, \linb: locations $r/D=1.4$, $y^+=15$.}
\label{fig:RST_bud_r2}
\end{figure}

The $rr$-component of the Reynolds stress tensor has, compared to the other components, much more significant terms in in its budget. Figure \ref{fig:RST_bud_r2} shows that the dominant terms at the wall are viscous diffusion (gaining stress) and turbulent dissipation (loosing stress). Farther away from the wall ($y^+=15$) pressure strain gains most stress. Its maximum is around $r/D \approx 0.8$. At this location, the stress is likewise decreased by turbulent diffusion.

\begin{figure}
\captionsetup[subfigure]{labelformat=empty}
\centering
\subfloat[]{\includegraphics[width=0.32\textwidth]{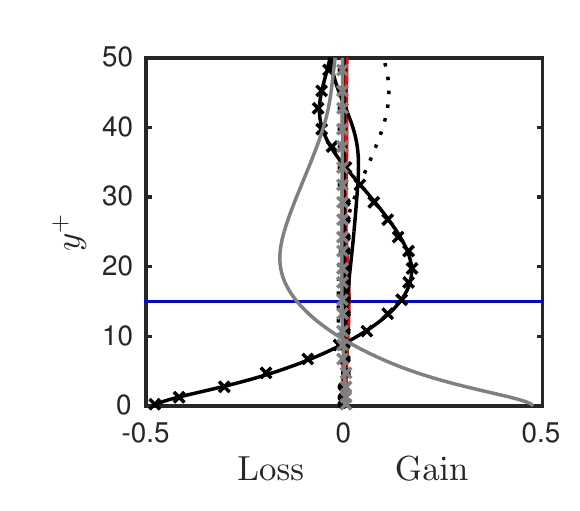}}
\subfloat[]{\includegraphics[width=0.32\textwidth]{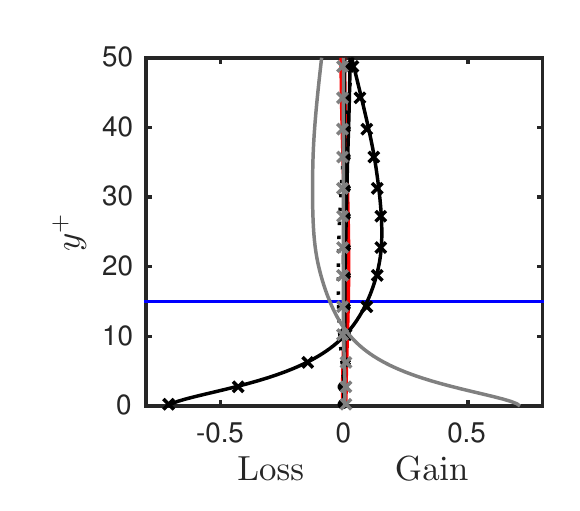}}\\[-10mm]

\subfloat[]{\includegraphics[width=0.32\textwidth]{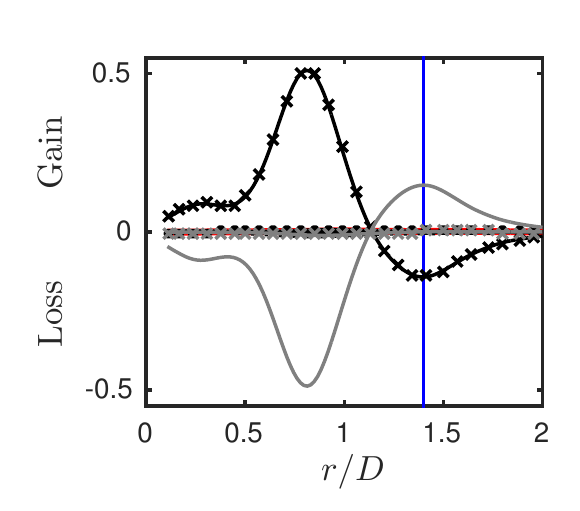}}
\subfloat[]{\includegraphics[width=0.32\textwidth]{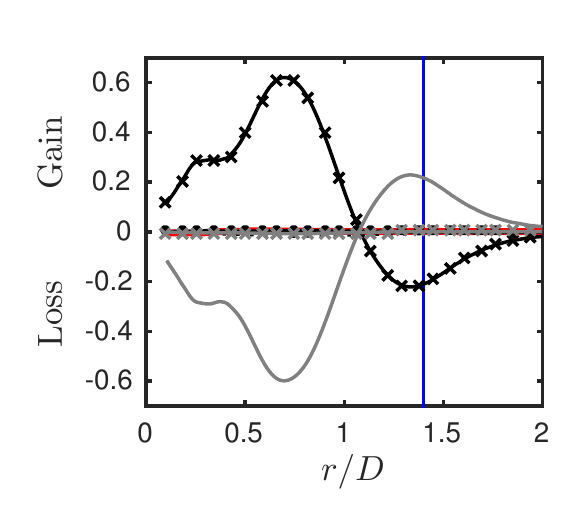}}\\[-8mm]
\caption{Budget of the $yr$-component $\ol{\rho v'' u_r''}$. Left column: \#5 ($\Rey=3300$), right colomn: \#6 ($\Rey=8000$), first row: $r/D=1.4$, second row: $y^+=15$. \budGGW: $IB$, \budC: $C$, \budPR: $PR$, \budTD: $TD$, \budVD: $VD$, \budM: $M$, \budPS: $PS$, \budDS: $DS$, \linb: locations $r/D=1.4$, $y^+=15$.}
\label{fig:RST_bud_yr}
\end{figure}

The budget of $\ol{\rho v'' u_r''}$ in figure \ref{fig:RST_bud_yr} contains two main terms: pressure strain and turbulent diffusion. The first one gains stress at and nearby the wall and then turns negative at higher dimensionless wall distances. In the boundary layer, the turbulent diffusion behaves opposite. At higher values of $y^+$, around 35, also production becomes more important in the case of $\Rey=3300$. The radial distance of $r/D=1.4$ is located in a local extremum of the two main terms. Another, stronger extremum with opposite signs occurs at $r/D \approx 0.8$.

\section{Conclusions}

In order to close a knowledge gap that presently exists in literature, we performed direct numerical simulations of compressible sub- and supersonic impinging jets and analysed their statistics. The main conclusions are as follows:
\begin{itemize}
	\item The influence of the Mach number on the velocity boundary layer is comparably low. The thermal boundary layer in contrary is stronger effected by the choice of $\Mac$.
	\item The heating of the impinging plate and the ambient fluid effects the mean dimensionless temperature and velocity profiles. The radial dimensionless velocity increases due to the heating.
	\item Reynolds analogies were tested for the subsonic compressible cases. The Nusselt number approximations according to the Reynolds and the Chilton Colburn analogies deliver useful values, if the distance to the jet axis is lager than one diameter. In the stagnation point region, both relations deliver big errors and cannot be recommended.
	\item The generalised Reynolds analogy (GRA) was applied and compared with the Crocco-Busemann relation. Both approximations relate the mean temperature field to the mean velocity field with inaccurate results. Only the GRA can predict the temperature field with good precision, in the area where the radial acceleration is zero ($r/D = 0.8$). Since the wall jet exhibits in most of the domain different flow conditions, that do not meet the assumed quasi-one-dimensional flow (canonical compressible wall-bounded turbulent flows), the relation cannot be applied reliably to the impinging jet.
	\item The relation between the fluctuating temperature and velocity according to the GRA is not applicable at all. The reason for this lies in the change of sign of the term $\ol{\rk{\rho v}' u'}$, which creates singularities in the predicted temperature fluctuations.
	\item Pressure fluctuations are not unimportant compared to the density fluctuation in the entire region where the flow is influenced by the impingement. This is different to the channel flow. Farther downstream however, the importance of the pressure fluctuations decreases as the influence of the impingement vanishes. As a result, the impingement region needs to be treated differently from the channel flow with regard on turbulence models.
	\item The linear relation between thermodynamic fluctuations of entropy, density and temperature as suggested by \cite{LechnerSesterhenn2001} can be confirmed for the entire wall jet.
	\item The existence of an area with higher local Nusselt number could be ascribed to the wall-normal heat flux, that transports hot fluid away from the wall at the radial distance of $1 \lesssim r/D \lesssim 2$ from the stagnation point. This is an important aspect for the increase of heat transfer efficiency, since this effect can be used and enhanced with pulsating inlet conditions. The phenomenon is stronger distinct at lower Reynolds numbers.
	\item The main components of the Reynolds stress tensor could be conciliated with the primary and secondary vortex rings of the wall jet. The budget terms are given in order to allow the improvement of RANS and LES models.\\
\end{itemize}

The simulations were performed on the national supercomputers Cray XE6 (Hermit) and Cray XC40 (Hornet, Hazelhen) at the High Performance Computing Center Stuttgart (HLRS) under the grant numbers GCS-NOIJ/12993 and GCS-ARSI/44027.

The authors gratefully acknowledge support by the Deutsche Forschungsgemeinschaft (DFG) as part of the collaborative research center SFB 1029 "Substantial efficiency increase in gas turbines through direct use of coupled unsteady combustion and flow dynamics".

\clearpage

\appendix
\section{Reynolds stress transport equations in cylindrical coordinates}\label{appA}

\subsubsection*{Convection}

\begin{align}
	C_{\theta\theta} &= -\left[\frac{1}{r}\pd{}{\theta}\left(\wt{u_{\theta}}\ol{\rho}\wt{u_{\theta}'' u_{\theta}''}\right)	+\pd{}{y}\left(\wt{v}\ol{\rho}\wt{u_{\theta}'' u_{\theta}''}\right)+\pd{}{r}\left(\wt{u_r}\ol{\rho}\wt{u_{\theta}'' u_{\theta}''}\right) \right.\nn\\
	&\left.+ \frac{\ol{\rho}}{r}\left(2\wt{u_r'' u_{\theta}''}\wt{u_{\theta}}+\wt{u_{\theta}'' u_{\theta}''}\wt{u_r}\right)\right] \nn\\
	C_{yy} &=-\left[\frac{1}{r}\pd{}{\theta}\left(\wt{u_{\theta}}\ol{\rho}\wt{v'' v''}\right)	+\pd{}{y}\left(\wt{v}\ol{\rho}\wt{v'' v''}\right)+\pd{}{r}\left(\wt{u_r}\ol{\rho}\wt{v'' v''}\right)+\frac{\ol{\rho}}{r}\wt{v'' v''}\wt{u_r}\right] \nn\\
	C_{yr} &= -\left[\frac{1}{r}\pd{}{\theta}\left(\wt{u_{\theta}}\ol{\rho}\wt{v'' u_r''}\right)	+\pd{}{y}\left(\wt{v}\ol{\rho}\wt{v'' u_r''}\right)+\pd{}{r}\left(\wt{u_r}\ol{\rho}\wt{v'' u_r''}\right) \right.\nn\\
	&\left.+\frac{\ol{\rho}}{r}\left(\wt{v'' u_r''}\wt{u_r} -\wt{v'' u_{\theta}''}\wt{u_{\theta}}\right)\right]\nn\\
	C_{rr} &= -\left[\frac{1}{r}\pd{}{\theta}\left(\wt{u_{\theta}}\ol{\rho}\wt{u_r'' u_r''}\right)	+\pd{}{y}\left(\wt{v}\ol{\rho}\wt{u_r'' u_r''}\right)+\pd{}{r}\left(\wt{u_r}\ol{\rho}\wt{u_r'' u_r''}\right) \right.\nn\\
	&\left.+\frac{\ol{\rho}}{r}\left(\wt{u_r'' u_r''}\wt{u_r} -2\wt{u_{\theta}'' u_r''}\wt{u_{\theta}}\right)\right]\nn\\
\end{align}

\subsubsection*{Production}

\begin{align}
	PR_{\theta\theta} &= -2\ol{\rho}\left[\wt{u_{\theta}'' u_{\theta}''} \left( \frac{1}{r}\pd{\wt{u_{\theta}}}{\theta}+\frac{\wt{u_r}}{r}\right) + \wt{u_{\theta}'' v''}\pd{\wt{u_{\theta}}}{y} + \wt{u_{\theta}'' u_r''}\pd{\wt{u_{\theta}}}{r}	\right] \nn\\
	&\left. + \wt{u_r'' u_{\theta}''}\left(\frac{1}{r}\pd{\wt{u_{\theta}}}{\theta}+\frac{\wt{u_r}}{r}\right) + \wt{u_r'' v''}\pd{\wt{u_{\theta}}}{y} + \wt{u_r'' u_{r}''}\pd{\wt{u_{\theta}}}{r}\right] \nn\\
	PR_{yy} &= -2\ol{\rho}\left[\wt{v'' u_{\theta}''}\frac{1}{r}\pd{\wt{v}}{\theta} +	\wt{v'' v''}\pd{\wt{v}}{y}
	+\wt{v'' u_r''}\pd{\wt{v}}{r}\right] \nn\\
	PR_{yr} &= -\ol{\rho}\left[\wt{v'' u_{\theta}''}\left(\frac{1}{r}\pd{\wt{u_r}}{\theta}-\frac{\wt{u_{\theta}}}{r}\right) +\wt{v'' v''}\pd{\wt{u_r}}{y} +\wt{v'' u_r''}\pd{\wt{u_r}}{r} + \wt{u_r'' u_{\theta}''}\frac{1}{r}\pd{\wt{v}}{\theta} \right. \nn\\
	&\left. + \wt{u_r'' v''}\pd{\wt{v}}{y} + \wt{u_r'' u_{\theta}''}\pd{\wt{v}}{r} \right] \nn\\
	PR_{rr} &= -2\ol{\rho}\left[\wt{u_r'' u_{\theta}''}\left(\frac{1}{r}\pd{\wt{u_r}}{\theta}-\frac{\wt{u_{\theta}}}{r}\right)+\wt{u_r'' v''}\pd{\wt{u_r}}{y}+\wt{u_r'' u_r''}\pd{\wt{u_r}}{r}\right]
\end{align}

\subsubsection*{Turbulent Diffusion}

\begin{align}
	TD_{\theta\theta} &=\pd{\ol{\rho u_{\theta}'' u_{\theta}'' u_r''}}{r}+\frac{1}{r}\pd{\ol{\rho u_{\theta}'' u_{\theta}'' u_{\theta}''}}{\theta} + \pd{\ol{\rho u_{\theta}'' u_{\theta}'' v''}}{y} + 3 \frac{\ol{\rho u_r'' u_{\theta}'' u_{\theta}''}}{r} \nn\\
	&+2\left(\frac{1}{r} \pd{\ol{p' u_{\theta}'}}{\theta} + \frac{\ol{p' u_r'}}{r}\right)  \nn\\
	TD_{yy} &=\pd{\ol{\rho v'' v'' u_r''} }{r} + \frac{1}{r}\pd{\ol{\rho v'' v'' u_{\theta}''} }{\theta}  + \pd{\ol{\rho v'' v'' v''}}{y} + \frac{\ol{\rho v'' v'' u_r''} }{r} + 2\pd{\ol{p'v'}}{y} \nn\\
	TD_{yr} &=\pd{\ol{\rho v'' u_r'' u_r''}}{r} + \frac{1}{r}\pd{\ol{\rho v'' u_r'' u_{\theta}''}}{\theta} + \pd{\ol{\rho v'' u_r'' v''}}{y} + \frac{\ol{\rho v'' u_r'' u_r''} -\ol{\rho v'' u_{\theta}'' u_{\theta}''} }{r} \nn\\
	&+ \pd{\ol{p'v'}}{r}+ \pd{\ol{p'u_r'}}{y} \nn\\
	TD_{rr} &=\pd{\ol{\rho u_r'' u_r'' u_r''}}{r} + \frac{1}{r} \pd{\ol{\rho u_r'' u_r'' u_{\theta}''}}{\theta} + \pd{\ol{\rho u_r'' u_r'' v''}}{y} +\frac{\ol{\rho u_r'' u_r'' u_r''}-2\ol{\rho u_r'' u_{\theta}'' u_{\theta}''}}{r} \nn\\
	&+ 2\pd{\ol{p'u_r'}}{r} \nn\\
\end{align}

\subsubsection*{Viscous Diffusion}

\begin{align}
	VD_{\theta\theta} &=2 \left( \pd{\ol{u_{\theta}' \tau_{\theta r}'}}{r}+\frac{1}{r}\pd{\ol{u_{\theta}' \tau_{\theta \theta}'}}{\theta} + \pd{\ol{u_{\theta}' \tau_{\theta y}'}}{y} + \frac{\ol{u_r' \tau_{\theta \theta}'}+2\ol{u_{\theta}' \tau_{r \theta}'}}{r} \right) \nn\\
	VD_{yy} &=2\left(\pd{\ol{v' \tau_{yr}'}}{r} + \frac{1}{r}\pd{\ol{v' \tau_{y\theta}'}}{\theta}  + \pd{\ol{v' \tau_{yy}'}}{y} + \frac{\ol{v' \tau_{yr}'}}{r} \right) \nn\\
	VD_{yr} &=\pd{\ol{v' \tau_{rr}'}}{r} + \frac{1}{r}\pd{\ol{v' \tau_{r\theta}'}}{\theta} + \pd{\ol{v' \tau_{ry}'}}{y} + \frac{\ol{v' \tau_{rr}'}-\ol{v' \tau_{\theta \theta}'}}{r} + 	\pd{\ol{u_r' \tau_{yr}'}}{r} \nn\\
	&+ \frac{1}{r}\pd{\ol{u_r' \tau_{y\theta}'}}{\theta} + \pd{\ol{u_r' \tau_{yy}'}}{y} +\frac{\ol{u_r' \tau_{yr}'}-\ol{u_{\theta}' \tau_{y\theta}'}}{r}	 \nn\\
	VD_{rr} &=2\left(	\pd{\ol{u_r' \tau_{rr}'}}{r} + \frac{1}{r} \pd{\ol{u_r' \tau_{r\theta}'}}{\theta} + \pd{\ol{u_r' \tau_{ry}'}}{y} +\frac{\ol{u_r' \tau_{rr}'}-\ol{u_{\theta}' \tau_{r\theta}'}-\ol{u_r' \tau_{\theta\theta}'}}{r} \right)
\end{align}

\subsubsection*{Mass-Flux Variation}

\begin{align}
	M_{\theta\theta} &= 2 \ol{u_{\theta}''}\left(\pd{\ol{\tau_{\theta r}}}{r} + \frac{1}{r}\pd{\ol{\tau_{\theta \theta}}}{\theta} + \pd{\ol{\tau_{\theta y}}}{y} + \frac{\ol{\tau_{r \theta}}+\ol{\tau_{\theta r}}}{r} -\frac{1}{r}\pd{\ol{p}}{\theta}\right) \nn\\
	M_{yy} &= 2\ol{v''}\left(\pd{\ol{\tau_{yr}}}{r}+\frac{1}{r}\pd{\ol{\tau_{y\theta}}}{\theta} + \pd{\ol{\tau_{yy}}}{y} +\frac{\ol{\tau_{yr}}}{r} -\pd{\ol{p}}{y}\right)\nn\\
	M_{yr} &= \ol{v''}\left(\pd{\ol{\tau_{rr}}}{r}+\frac{1}{r}\pd{\ol{\tau_{r\theta}}}{\theta} +\pd{\ol{\tau_{ry}}}{y} + \frac{\ol{\tau_{rr}} - \ol{\tau_{\theta\theta}}}{r} -\pd{\ol{p}}{r}\right) \nn\\
	&+ \ol{u_r''}\left(\pd{\ol{\tau_{yr}}}{r}+\frac{1}{r}\pd{\ol{\tau_{y\theta}}}{\theta} + \pd{\ol{\tau_{yy}}}{y} +\frac{\ol{\tau_{yr}}}{r}	-\pd{\ol{p}}{y}\right)\nn\\
	M_{rr} &= 2\ol{u_r''}\left(\pd{\ol{\tau_{rr}}}{r}+\frac{1}{r}\pd{\ol{\tau_{r\theta}}}{\theta} +\pd{\ol{\tau_{ry}}}{y} + \frac{\ol{\tau_{rr}} - \ol{\tau_{\theta\theta}}}{r} -\pd{\ol{p}}{r}\right)
\end{align}

\subsubsection*{Pressure Strain}

\begin{align}
	PS_{\theta \theta} &= \frac{2}{r}\ol{p'\left(\pd{u_{\theta}'}{\theta} + u_r'\right)} \nn\\
	PS_{rr} &= 2\ol{p' \pd{u_{r}'}{r}} \nn\\	
	PS_{ry} &= \ol{p'\left(\pd{u_{r}'}{y}+\pd{v'}{r}\right)} \nn\\	
	PS_{yy} &= 2\ol{p'\pd{v'}{y}}
\end{align}


\subsubsection*{Turbulent Dissipation}

\begin{align}
	DS_{\theta \theta} &=-2\left[\ol{\tau_{\theta \theta}'\left(\frac{1}{r}\pd{u_{\theta}'}{\theta} + \frac{u_r'}{r}\right)+\tau_{\theta y}' \pd{u_{\theta}'}{y}+\tau_{\theta r}' \pd{u_{\theta}'}{r}}\right] \nn\\
	DS_{yy} &=-2\left[\ol{\tau_{y\theta}' \frac{1}{r}\pd{v'}{\theta} + \tau_{yy}' \pd{v'}{y} + \tau_{yr}' \pd{v'}{r}}\right] \nn\\
	DS_{yr} &=-\left[\ol{\tau_{y\theta}' \left(\frac{1}{r}\pd{u_{r}'}{\theta}-\frac{u_{\theta}'}{r}\right) +\tau_{yy}' \pd{u_{r}'}{y} +\tau_{yr}' \pd{u_{r}'}{r}} \right. \nn\\
	&\left. + \ol{\tau_{r\theta}' \frac{1}{r}\pd{v'}{\theta} + \tau_{ry}' \pd{v'}{y} + \tau_{rr}' \pd{v'}{r}}\right] \nn\\
		DS_{rr} &=-2\left[\ol{\tau_{r\theta}' \left(\frac{1}{r}\pd{u_{r}'}{\theta}-\frac{u_{\theta}'}{r}\right) +\tau_{ry}' \pd{u_{r}'}{y} +\tau_{rr}' \pd{u_{r}'}{r}}\right]
\end{align}

\bibliographystyle{jfm}
\bibliography{0_literatur_impjet}

\end{document}